\documentstyle[epsfig]{mn}
\input{epsf}

\onecolumn
 
\renewcommand{\d}{{\rm d}}
\newcommand{\beq}{\begin{equation}}
\newcommand{\eeq}{\end{equation}}
\newcommand{\beqa}{\begin{eqnarray}} 
\newcommand{\eeqa}{\end{eqnarray}}
\newcommand{\bea}{\begin{array}} 
\newcommand{\ea}{\end{array}} 
\newcommand{\lag}{\langle}
\newcommand{\rag}{\rangle}
\newcommand{\Om}{\Omega_{\rm m}}
\newcommand{\Ol}{\Omega_{\Lambda}}
\newcommand{\De}{{\cal D}}

\newcommand{\Map}{M_{\rm ap}}
\newcommand{\bx}{{\bf x}}
\newcommand{\bk}{{\bf k}}
\newcommand{\kpar}{k_{\parallel}}
\newcommand{\kperp}{\bk_{\perp}}
\newcommand{\kperpDt}{k_{\perp}\De\theta_s}

\newcommand{\wt}{\tilde{w}}
\newcommand{\gammai}{\gamma_i}
\newcommand{\gammais}{\gamma_{i{\rm s}}}
\newcommand{\Mapone}{M_{\rm ap1}}
\newcommand{\Maptwo}{M_{\rm ap2}}
\newcommand{\gamisone}{\gamma_{i{\rm s}1}}
\newcommand{\gamistwo}{\gamma_{i{\rm s}2}}
\newcommand{\cM}{{\cal M}}
\newcommand{\cH}{{\cal H}}
\newcommand{\cS}{{\cal S}}
\newcommand{\cR}{{\cal R}}
\newcommand{\cT}{{\cal T}}

\def\e{\epsilon}

\title[Cross-correlating Weak Lensing Surveys]
{On Cross-correlating Weak Lensing Surveys}
\author[Munshi \& Valageas]
{Dipak Munshi$^{1,2}$, Patrick Valageas$^{3}$\\
$^{1}$Institute of Astronomy, Madingley Road,
Cambridge, CB3 OHA, United Kingdom\\
$^{2}$Astrophysics Group, Cavendish Laboratory, Madingley Road, 
Cambridge CB3 OHE, United Kingdom\\
$^{3}$Service de Physique Th\'eorique, 
CEA Saclay, 91191 Gif-sur-Yvette, France \\
}

\begin{document}
\maketitle

\begin{abstract}
The present generation of weak lensing surveys will be superseded by surveys 
run from space with much better sky coverage and high level of signal to
noise ratio, such as SNAP. However, removal of any systematics or noise 
will remain a major cause of concern for any weak lensing survey. One of 
the best ways of spotting any undetected source of systematic noise is to 
compare surveys which probe the same part of the sky. In this paper we 
study various measures which are useful in cross correlating weak lensing 
surveys with diverse survey strategies. Using two different statistics - 
the shear components and the aperture mass - we construct a class of 
estimators which encode such cross-correlations. These techniques will 
also be useful in studies where the entire source population from a specific 
survey can be divided into various redshift bins to study cross correlations 
among them. We perform a detailed study of the angular size dependence and 
redshift dependence of these observables and of their sensitivity to the
background cosmology. We find that one-point and two-point statistics provide 
complementary tools which allow one to constrain cosmological parameters and
to obtain a simple estimate of the noise of the survey.     
\end{abstract}

\begin{keywords}
Cosmology: theory -- gravitational lensing -- large-scale structure 
of Universe -- Methods: analytical, statistical, numerical
\end{keywords}

\section{Introduction}

Detection of weak lensing signals by observational teams 
(e.g., Bacon, Refregier \& Ellis, 2000, Hoekstra et al., 2002, 
Van Waerbeke et al., 2000, and Van Waerbeke et al., 2002) has led to a new 
avenue not only in constraining the background dynamics of the universe but 
in probing the nature of dark matter and dark energy as well. To do so one
compares observations with theoretical results obtained from simulations
or analytical methods.


Numerical simulations of weak lensing typically employ
ray-tracing techniques as well as line of sight integration of cosmic shear
(e.g., Schneider \& Weiss, 1988, Jarosszn'ski et al., 1990, 
Wambsganns, Cen \& Ostriker, 1998, Van Waerbeke, Bernardeau \& Mellier, 1999, 
and Jain, Seljak \& White, 2000,Couchman, Barber \& Thomas (1999)) and
provide valuable insights.


On the other hand, analytical techniques to study weak lensing include 
perturbative calculations at large angular scales (e.g., Villumsen, 1996, 
Stebbins, 1996, Bernardeau et al., 1997, Jain \& Seljak, 1997, Kaiser, 1998, 
Van Waerbeke, Bernardeau \& Mellier, 1999, and Schneider et al., 1998) and 
techniques based on the hierarchical {\em ansatz} at small angular scales
(e.g., Fry 1984, Schaeffer 1984, Bernardeau \& Schaeffer 1992, 
Szapudi \& Szalay 1993, 1997, Munshi, Melott \& Coles 1999). Ingredients 
for such calculations include Peacock \& Dodds (1996)'s prescription 
(see Peacock \& Smith (2000) for a more recent fit) for the evolution of the 
power spectrum or equivalently the two-point correlation function. Recent 
studies have shown an excellent agreement between analytical results and 
numerical simulations of weak lensing effects (Valageas 2000a \& b; 
Munshi \& Jain 2000 \& 2001; Munshi 2000; Bernardeau \& Valageas 2000; 
Valageas, Barber \& Munshi 2004; Barber, Munshi \& Valageas 2004; 
Munshi, Valageas \& Barber 2004).


However, to the weak lensing effects themselves one must add various sources
of noise, such as the intrinsic ellipticity distribution of galaxies, 
shot-noise due to the discreet nature of the source galaxies and finite 
volume effects due to finite survey size. There is a  need to incorporate all 
these in a consistent and systematic way in any realistic design of survey 
strategy. A detailed formalism to tackle such issues was developed by 
Munshi \& Coles (2002) following Schneider et al. (1998). Recently 
Valageas, Munshi \& Barber (2004) have used such techniques and have 
incorporated the redshift distribution of sources to model errors in
future surveys such as SNAP.


In the near future we expect to see many weak lensing surveys probing
overlapping areas of the sky with different survey strategies.
These ground based observations will finally be superseded
by surveys run from space with much better sky coverage such as SNAP.
Nevertheless, removal of systematics and control of the noise from individual 
surveys will remain a major cause of concern. It is therefore of
utmost importance that any undetected systematics be spotted by comparing 
different surveys which probe the same regions of the sky. Such a cross 
comparison will finally lead to the construction of large maps by combining
smaller maps in an optimum way.


In our present study we focus on cross correlations among various surveys or 
different redshift bins in the same survey, with partial or complete overlap 
on their sky coverage. Using two different classes of statistics such as the 
smoothed shear components and the aperture mass, we construct estimators which 
encode such correlations. Based on analytical results from 
Munshi \& Coles (2002) we generalize a recent study by 
Valageas, Munshi \& Barber (2004) to take into account multiple surveys.
We display the power of such techniques using SNAP class surveys where the 
source population can be subdivided into various redshift bins to study cross 
correlation among subsamples. We study the angular size dependence and 
redshift dependence of these classes of statistics. We also describe the
dependence on background cosmological parameters.

This paper is organized as follows: in section~2, we describe our notations
regarding weak lensing observables in general and various filters used to 
smooth the data. In section~3, we introduce weak lensing estimators which 
take into account various realistic sources of noise. Specific survey 
geometries based on SNAP class experiments are introduced in section~4 and we 
compute the noisy cumulant correlators of the aperture mass and of the shear 
components. Section~5 is devoted to a discussion of our results.

\section{Notations and Formalism}
\label{notations}

Let us first recall our notations. Weak-lensing effects can be expressed 
in terms of the convergence along the line-of-sight towards the direction 
${\vec \vartheta}$ on the sky up to the redshift $z_s$ of the source, 
$\kappa({\vec \vartheta},z_s)$, given by (e.g., Bernardeau et al. 1997; 
Kaiser 1998):
\beq
\kappa({\vec \vartheta},z_s) = \frac{3\Om}{2} \int_0^{\chi_s} \d\chi \; 
w(\chi,\chi_s) \; \delta(\chi,\De{\vec \vartheta}) , \;\;\; \mbox{with} 
\; \; \; w(\chi,\chi_s) = \frac{H_0^2}{c^2} \; \frac{\De(\chi) 
\De(\chi_s-\chi)}{\De(\chi_s)} \; (1+z) ,
\label{w}
\eeq
where $z$ corresponds to the radial distance $\chi$ and $\De$ is the angular 
distance. Here and in the following we use the Born approximation which is
well-suited to weak-lensing studies: the fluctuations of the gravitational
potential are computed along the unperturbed trajectory of the photon 
(Kaiser 1992). Thus the convergence $\kappa({\vec \vartheta},z_s)$ is 
merely the projection of the local density contrast $\delta$ along the 
line-of-sight. Therefore, weak lensing observations 
allow us to measure the projected density field $\kappa$ 
on the sky (note that by looking at sources located at different redshifts 
one may also probe the radial direction). In practice the sources have a broad
redshift distribution which needs to be taken into account. Thus, the quantity
of interest is actually:
\beq
\kappa({\vec \vartheta}) = \int_0^{\infty}\d z_s \; n(z_s) 
\kappa({\vec \vartheta},z_s) , \;\;\; \mbox{with} \; \int\d z_s \; n(z_s)=1 ,
\label{kappanz}
\eeq
where $n(z_s)$ is the mean redshift distribution of the sources 
(e.g. galaxies) normalized to unity. From eq.(\ref{w}), the convergence
$\kappa$ associated with a specific survey also reads:
\beq
\kappa({\vec \vartheta}) = \int_0^{\chi_{\rm max}} \d\chi \; \wt(\chi) \; 
\delta(\chi,\De{\vec \vartheta}) , \;\;\; \mbox{with} \; \wt(\chi) = 
\frac{3\Om}{2} \int_z^{z_{\rm max}} \d z_s \; n(z_s) \; w(\chi,\chi_s) ,
\label{kappa}
\eeq
where $z_{\rm max}$ is the depth of the survey (i.e. $n(z_s)=0$ for 
$z_s>z_{\rm max}$).
By working with eq.(\ref{kappa}) we neglect the discrete effects due to the 
finite number of galaxies. They can be obtained by taking into account the 
discrete nature of the distribution $n(z_s)$. This gives corrections of 
order $1/N$ to higher-order moments of weak-lensing observables, where $N$ 
is the number of galaxies within the circular field of interest. In practice 
$N$ is much larger than unity (for a circular window of radius 1 arcmin we 
expect $N \ga 100$ for the SNAP mission) therefore in this paper we shall 
work with eq.(\ref{kappa}).

Thus, in order to take into account the redshift distribution of sources we 
simply need to replace $3\Om/2 w(\chi,\chi_s)$ in eq.(\ref{w}) by $\wt(\chi)$.
Therefore, all the results of Munshi et al. (2004) remain valid.
Then, usual weak-lensing observables can be written as the angular average 
of $\kappa({\vec \vartheta})$ with some filter $U$:
\beq
X= \int\d{\vec \vartheta}\;U_X({\vec \vartheta})\;\kappa({\vec \vartheta}).
\label{X}
\eeq
For instance, the filters associated with the smoothed convergence $\kappa_s$,
the smoothed shear $\gamma_s$ and the aperture-mass $\Map$ are 
(Munshi et al. 2004):
\beq
U_{\kappa_s}= \frac{\Theta(\vartheta<\theta_s)}{\pi\theta_s^2} , \;\;\; 
U_{\gamma_s} = - \frac{\Theta(\vartheta>\theta_s)}{\pi\vartheta^2}
\; e^{i2\alpha}  \;\;\; \mbox{and} \;\;\; U_{\Map} = 
\frac{\Theta(\vartheta<\theta_s)}{\pi\theta_s^2}
\; 9 \left(1-\frac{\vartheta^2}{\theta_s^2}\right) 
\left(\frac{1}{3} - \frac{\vartheta^2}{\theta_s^2}\right) ,
\label{UMap}
\eeq
where $\Theta$ are Heaviside functions with obvious notations and $\alpha$ 
is the polar angle of the vector ${\vec \vartheta}$. The angular radius 
$\theta_s$ gives the angular scale probed by these smoothed observables. Note
that the smoothed shear $\gamma_s$ depends on the matter located outside of
the cone of radius $\theta_s$. However, in practice one directly measures
the shear $\gamma({\vec \vartheta})$ on the direction ${\vec \vartheta}$
(from the ellipticity of a galaxy) rather than the convergence $\kappa$ and 
$\gamma_s$ is simply the mean shear within the radius $\theta_s$. For $\Map$ 
we shall use in this paper the filter (\ref{UMap}), as in Schneider (1996), 
but one could also use any compensated filter with radial symmetry.

As described in Munshi et al. (2004), the cumulants of $X$ can be
written as:
\beq
\lag X^p\rag_c = \int_0^{\infty} \prod_{i=1}^{p} \d\chi_i \; \wt(\chi_i) 
\int \prod_{j=1}^{p} \d{\vec \vartheta}_j \; U_X({\vec \vartheta}_j) \;\; 
\xi_p\left( \bea{l} \chi_1 \\ \De_1 {\vec \vartheta}_1 \ea ,
\bea{l} \chi_2 \\ \De_2 {\vec \vartheta}_2 \ea , \dots , 
\bea{l} \chi_p \\ \De_p {\vec \vartheta}_p \ea \right) ,
\label{cumXr}
\eeq
or equivalently we can write:
\beq
\lag X^p \rag_c = \int_0^{\infty} \prod_{i=1}^{p} \d\chi_i \; \wt(\chi_i)
\int \prod_{j=1}^{p} \d\bk_j \; W_X(\bk_{\perp j} \De_j \theta_s) \;\;
\left( \prod_{l=1}^{p} e^{i k_{\parallel l} \chi_l} \right) 
\;\; \lag \delta(\bk_1) \dots \delta(\bk_p) \rag_c .
\label{cumXk}
\eeq
We note $\lag .. \rag$ the average over different realizations of the
density field, $\xi_p$ is the real-space $p-$point correlation function of 
the density field $\xi_p(\bx_1,..,\bx_p)= \lag \delta(\bx_1) .. 
\delta(\bx_p)\rag_c$, $\kpar$ is the component of $\bk$ parallel to the 
line-of-sight, $\kperp$ is the two-dimensional vector formed by the 
components of $\bk$ perpendicular to the line-of-sight and 
$W_X(\kperp\De\theta_s)$ is the Fourier transform of the window $U_X$:
\beq
W_X(\kperp\De\theta_s) = \int\d{\vec \vartheta} \; U_X({\vec \vartheta}) 
\; e^{i \kperp.\De{\vec \vartheta}} .
\label{WX}
\eeq
In particular, for the smoothed convergence $\kappa_s$, the smoothed shear
$\gamma_s$ and the aperture-mass $\Map$ we have (Munshi et al. 2004):
\beq
W_{\kappa_s}(\kperp\De\theta_s) = \frac{2 J_1(\kperpDt)}{\kperpDt} , \;\;\; 
W_{\gamma_s}(\kperp\De\theta_s) = \frac{2 J_1(\kperpDt)}{\kperpDt} 
\; e^{i 2\phi} \;\;\; \mbox {and} \;\;\;
W_{\Map}(\kperp\De\theta_s) =  \frac{24 J_4(\kperpDt)}{(\kperpDt)^2} ,
\label{WMap}
\eeq
where $\phi$ is the polar angle of $\kperp$ and $J_{\nu}$ are Bessel functions
of the first kind. The real-space expression (\ref{cumXr}) is well-suited
to models which give an analytic expression for the correlations $\xi_p$,
like the minimal tree-model (Valageas 2000b; Bernardeau \& Valageas 2002; 
Barber et al. 2004) while the Fourier-space expression (\ref{cumXk}) is 
convenient for models which give a simple expression for the correlations
$\lag \delta(\bk_1) .. \delta(\bk_p)\rag_c$, like the stellar model 
(Valageas et al. 2004; Barber et al. 2004). In the present study we shall also
be interested in two-point cumulants such as:
\beq
\lag X_1^p X_2^q \rag_c = \int_0^{\infty} \d\chi \; (2\pi)^{p+q-1} 
\; \wt_1^p \wt_2^q \int \prod_{j=1}^{p+q} \d\bk_{\perp j} 
\; W_X(\bk_{\perp j} \De \theta_s) \; 
\lag \delta(\bk_{\perp 1}) \dots \delta(\bk_{\perp p+q}) \rag_c ,
\label{cum2Xk}
\eeq
which describe the cross-correlations between two surveys or subsamples.
Here the subscripts ``1,2'' refer to the two surveys we cross-correlate.
In eq.(\ref{cum2Xk}) we used the fact that the correlation length is much 
smaller than cosmological scales and the Dirac factor 
$\delta_D(k_{\parallel 1}+\dots+k_{\parallel p+q})$ has been factorized out
of $\lag \delta(\bk_{\perp 1}) \dots \delta(\bk_{\perp p+q}) \rag_c$.
The two-point correlators are similar to projected cumulant correlators studied
in previous works (Szapudi \& Szalay 1993; Munshi et al. 1999 ). However, 
unlike those
projected cumulant correlators these objects are defined here in the zero 
angular separation limit where only the projection along the radial direction 
is different for the two surveys. Note that for observables like the 
aperture-mass which are very localized (long wavelengths are damped by the
compensated filter) it is expected that projected cumulant correlators will 
only have non-negligible values if the angular separation is not much larger 
than the angular radius of the window.

From the one-point cumulants (\ref{cumXr})-(\ref{cumXk}) it is convenient
to define the parameters $S_p$ and the coefficients $t_p$:
\beq
S_p = \frac{\lag X^p\rag_c}{\lag X^2\rag_c^{p-1}} , \hspace{1cm}
t_p = \frac{\lag X_1^p\rag_c}{\lag X_2^p\rag_c} .
\label{Sp}
\eeq
The quantities $S_p$ apply to one subsample while $t_p$ apply to two
subsamples. The coefficients $S_p$ of weak lensing observables are closely 
related to the usual parameters $S_p$ which describe the departures of the 
underlying 3-d density field from Gaussianity (e.g., Valageas et al. 2004, 
Munshi et al. 2004). In particular, for the case of the smoothed convergence 
$\kappa_s$ we have $S_p^{{\hat \kappa}_s}\simeq S_p^{\delta}$ where 
${\hat \kappa}_s$ is the smoothed convergence normalized by a suitable 
factor and $\delta$ is the 3-d density contrast (e.g., Barber et al. 2004). 
The normalization of the coefficients $S_p^{\delta}$ is such that they have 
a finite limit in the quasi-linear regime and show at most weak variations 
in the highly non-linear regime (but they exhibit a sharp change at the 
transition). Thus, from the one-point quantities $S_p$ we obtain the 
deviations 
from Gaussianity associated with each survey or subsample. Together with the 
second-order moments $\lag X^2\rag$ (i.e. variance) they can be used to 
measure cosmological parameters as well as the non-Gaussianities built by 
the non-linear gravitational dynamics. In this regard, the redshift
binning of the source population can also be very useful as it allows
one to constrain the time evolution of cosmological distances and growth 
factors through the ratios $t_p$. 
Obviously, the weak lensing effects associated with different redshift 
bins are correlated since their lines of sight probe the same density
fluctuations at low $z$ (where they are largest), for a fixed angular patch 
on the sky. In order to measure these cross-correlations we define the
cross-correlation coefficients $r_{pq}$ as:
\beq
r_{pq} = \frac{\lag X_1^p X_2^q \rag_c}{\lag X_1^{p+q}\rag_c^{p/(p+q)}
\lag X_2^{p+q}\rag_c^{q/(p+q)}} \; , \;\;\; \mbox{in particular}
\;\;\; r_{11} = \frac{\lag X_1 X_2 \rag_c}{\lag X_1^2\rag_c^{1/2}
\lag X_2^2\rag_c^{1/2}}  .
\label{rpq}
\eeq
The quantities $r_{pq}$ correspond to the two-point cumulants 
$\lag X_1^p X_2^q \rag_c$ normalized in such a way that most of the dependence
on cosmology and gravitational dynamics cancels out. Thus, if the two 
subsamples are highly correlated we have $r_{pq} \simeq 1$ while 
$r_{pq} \simeq 0$ if they are almost uncorrelated.

\begin{figure}
\protect\centerline{
\epsfysize = 2.25truein
\epsfbox[1 1 418 324]
{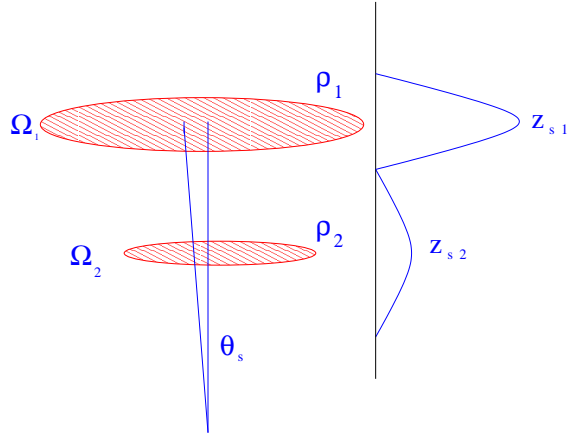}}
\caption{A schematic representation of cross-correlating two different
source populations represented by mean source redshifts
$z_{s1}$ and $z_{s2}$. Non-overlapping source distributions are considered. 
Survey areas are $\Omega_1$ and $\Omega_2$ respectively, the common
smoothing angular radius is $\theta_s$. The contributions to the scatter of
weak lensing estimators due to the galaxy intrinsic ellipticities of each
sample can be described by the quantities $\rho_1(\theta_s)$ and 
$\rho_2(\theta_s)$ defined in eq.(\ref{rho}).}
\label{survey}
\end{figure}

One often uses cumulant correlators to study the 
cross-correlations as a function of the separation angle $\theta_{12}$ 
between two different angular patches. In the numerical computations we
perform in this paper we shall restrict ourselves to the case of zero angular 
separation. Hence we focus on the correlations between two surveys which 
overlap on the sky but which have different redshift distributions of sources
(see Fig.~\ref{survey}). On the other hand, note that the signal would 
decrease for nonzero angular separations.

\section{Low-order estimators and their scatter}
\label{estimators}

The expressions obtained in the previous section describe weak lensing effects
due to the fluctuations of the matter density field. However, in order to
handle realistic data one needs to take into account the observational
noise. A first source of scatter is due to the finite size of the surveys. 
A second source of noise is merely due to the intrinsic ellipticity 
of galaxies, which cannot be avoided. As in Munshi \& Coles (2003) and
Valageas, Munshi \& Barber (2004), this
leads us to define the estimators $M_p$ for low-order moments:
\beq
M_p = \frac{(\pi \theta_s^2)^p} {(N)_p} \left [ \sum_{(i_1,
\dots, i_p)}^N Q_{i_1} \dots Q_{i_p} \; \e_{i_1} \dots \e_{i_p}
\right ], \;\;\; \mbox{with} \;\;\; 
(N)_p = N (N-1) .. (N-p+1) = \frac{N!}{(N-p)!} ,
\label{Mp}
\eeq
where $N$ is the number of galaxies in the patch of size $\pi
\theta_s^2$ and $p$ is the order of the moment. Here we used the fact that 
the aperture-mass defined from the convergence $\kappa$ by the compensated 
filter $U_{\Map}$ given in eq.(\ref{UMap}) can also be written as a function 
of the tangential shear $\gamma_{\rm t}$ as (Kaiser et al. 1994; 
Schneider 1996):
\beq
\Map= \int \d{\vec \vartheta} \; Q_{\Map}({\vec \vartheta}) \; 
\gamma_{\rm t}({\vec \vartheta}) \;\;\; \mbox{with} \;\;\;
Q_{\Map}({\vec \vartheta}) = \frac{\Theta(\vartheta<\theta_s)}{\pi\theta_s^2}
\; 6  \; \left(\frac{\vartheta}{\theta_s}\right)^2 
\left(1-\frac{\vartheta^2}{\theta_s^2}\right)  ,
\label{QMap}
\eeq
while for the smoothed shear components $\gammais$ (with $i=1,2$) we simply
have:
\beq
\gammais = \int \d{\vec \vartheta} \; Q_{\gammais}({\vec \vartheta}) \;
\gammai({\vec \vartheta}) \;\;\; \mbox{with} \;\;\;
Q_{\gammais}({\vec \vartheta}) = \frac{\Theta(\vartheta<\theta_s)}
{\pi\theta_s^2} .
\label{Qgammais}
\eeq
Thus, in eq.(\ref{Mp}) we wrote $Q_j=Q_X({\vec \vartheta}_j)$ and we use
the tangential ellipticity (for $\Map$) or the $i$-component of the ellipticity
(for $\gammais$). Indeed, in the case of weak lensing, $\kappa \ll 1$, 
the observed complex ellipticity $\epsilon$ is related to the shear $\gamma$ 
by: $\epsilon=\gamma+\epsilon_*$, where $\epsilon_*$ is the intrinsic 
ellipticity of the galaxy. Finally, the sum in eq.(\ref{Mp}) runs over 
all ordered sets of $p$ different galaxies among the $N$ galaxies enclosed 
in the angular radius $\theta_s$.

Then, neglecting any cross-correlations between the intrinsic ellipticities
of different galaxies and between the matter density field and the galaxy
intrinsic ellipticities, the estimators $M_p$ introduced in eq.(\ref{Mp})
are unbiased estimators of low-order moments of the aperture-mass or of the
shear components. Moreover, their dispersion 
$\sigma^2(M_p)=\lag M_p^2\rag -\lag M_p\rag^2$ only involves the variance
$\sigma_*^2=\lag|\epsilon_*|^2\rag$ of the galaxy intrinsic ellipticities
(even if the latter are not Gaussian). The latter enters the scatter 
$\sigma^2$ through the combination $\rho$ defined by:
\beq
\rho = \frac{2N\lag X^2\rag}{\sigma_*^2 G_X} \;\;\; \mbox{with} \;\;\;
G_X= \pi\theta_s^2 \int \d{\vec \vartheta} \; Q_X({\vec \vartheta})^2 ,
\;\;\; \mbox{whence} \;\;\; G_{\Map} = \frac{6}{5} , \;\;\; 
G_{\gammais} = 1 .
\label{rho}
\eeq
Here $N$ is the number of galaxies within the circular field of angular radius
$\theta_s$. The quantity $\rho$ measures the relative importance of the
galaxy intrinsic ellipticities in the signal. They can be neglected if 
$\rho\gg 1$. Thus, the mean of the estimator $M_2$ and its scatter are
given by (with $\lag X\rag=0$, see also Valageas et al. 2004):
\beq
\lag M_2 \rag = \lag X^2 \rag_c \;\;\; \mbox{and} \;\;\;
\sigma^2(M_2) = \lag X^4\rag_c + \lag X^2 \rag_c^2 \; 2 \;
\left[1+{1\over\rho}\right]^2 .
\label{M2}
\eeq
In order to obtain eq.(\ref{M2}) we have averaged i) over 
the intrinsic ellipticity distribution, ii) over the galaxy positions and 
iii) over the matter density field, assuming these three averaging procedures 
are uncorrelated. Any Gaussian white noise associated with the detector can 
be incorporated into these expressions by adding a relevant correction to 
$\sigma_*^2$ (whence to $\rho$). Note that the scatter of low-order estimators 
involves higher-order cumulants (up to twice the order of the moment one 
wishes to evaluate). Here we have neglected the terms due to the finite 
number of galaxies which scale as $1/N$ and are negligible for practical 
purposes ($N\gg 1$). The dispersion $\sigma^2$ obtained in eq.(\ref{M2})
takes into account two sources of noise: the galaxy intrinsic ellipticities
(the terms which depend on $\rho$) and the finite size of the survey
(the remaining terms).

The estimators $M_p$ defined in eq.(\ref{Mp}) correspond to a single
circular field of angular radius $\theta_s$ containing $N$ galaxies. In
practice, the size of the survey is much larger than $\theta_s$ and we can
average over $N_c$ cells on the sky. This yields the estimators $\cM_p$
defined by:
\beq
\cM_p = \frac{1}{N_c} \sum_{n=1}^{N_c} M_p^{(n)} , \;\;\; \mbox{whence}
\;\;\; \lag \cM_p \rag = \lag M_p \rag = \lag X^p \rag \;\;\; \mbox{and}
\;\;\; \sigma(\cM_p) = \frac{\sigma(M_p)}{\sqrt{N_c}} \;\;\; \mbox{with}
\;\;\; \sigma^2(\cM_p)=\lag \cM_p^2\rag -\lag \cM_p\rag^2 ,
\label{cMp}
\eeq
where $M_p^{(n)}$ is the estimator $M_p$ for the cell $n$ and we assumed 
that these cells are sufficiently well separated so as to be uncorrelated.

The estimators $M_p$ and $\cM_p$ provide a measure of the moments 
$\lag X^p\rag$ of weak lensing observables, which can be used to evaluate
the $S_p$ parameters defined in eq.(\ref{Sp}). However, as shown in 
Valageas et al. (2004), it is better to first consider cumulant-inspired
estimators $H_p$. Thus, for the aperture mass where we shall restrict ourselves
to third-order cumulants we define:
\beq
H_3 = M_3 - 3 \cM_2 M_1 \;\;\; \mbox{and} \;\;\; 
\cH_3 = \frac{1}{N_c} \sum_{n=1}^{N_c} H_3^{(n)} , \;\;\; \mbox{whence}
\;\;\; \lag \cH_3 \rag = \lag \Map^3 \rag_c ,
\label{H3}
\eeq
while for the shear components which are even quantities (so that all 
odd-order moments vanish) we need to go up to fourth order and we define:
\beq
H_4 = M_4 - 6 \cM_2 M_2 + 3 \cM_2^2 \;\;\; \mbox{and} \;\;\; 
\cH_4 = \frac{1}{N_c} \sum_{n=1}^{N_c} H_4^{(n)} , \;\;\; \mbox{whence}
\;\;\; \lag \cH_4 \rag = \lag \gammais^4 \rag_c .
\label{H4}
\eeq
The interest of $H_3$ and $H_4$ is that their scatter is smaller than for
$M_3$ and $M_4$, see Valageas et al. (2004). Besides they directly yield the
one-point cumulants (here we neglected higher-order terms over $1/N_c$).

As for galaxy surveys (Szapudi \& Szalay 1997) we can generalize the 
estimators (\ref{Mp}) to measure the cross-correlations between two surveys
(or two subsamples) by defining:
\beq
M_{pq} = \frac{(\pi\theta_{s1}^2)^p (\pi\theta_{s2}^2)^q}{(N_1)_p (N_2)_q} 
\left [ \sum_{(i_1,\dots,i_p)}^{N_1}
\sum_{(j_1,\dots, j_{q})}^{N_2}  Q_{i_1} \e_{i_1} \dots Q_{i_p} \e_{i_p} 
Q_{j_1} \e_{j_1} \dots  Q_{j_q} \e_{j_q} \right ].
\label{Mpq}
\eeq
Obviously, these estimators could be extended to $s-$point correlations
between $s$ surveys. In this paper we shall only consider the case 
$\theta_{s1}=\theta_{s2}$. Then, as in eq.(\ref{cMp}) we can define the
estimators $\cM_{pq}$ by averaging $M_{pq}$ over $N_c$ cells. Of course,
the one-point estimators can also be written $M_p=M_{p0}$ or $M_p=M_{0p}$
depending on which survey they apply to. Next, we can define the analogs of
the one-point estimators $H_p$ and $\cH_p$. For the aperture mass, there is
only one new cross-correlation at third order, $M_{21}$ (and its symmetric 
$M_{12}$), and we define:
\beq
H_{21} = M_{21} - 2 \cM_{11} M_{10} - \cM_{20} M_{01} \;\;\; \mbox{whence}
\;\;\; \lag \cH_{21} \rag = \lag \Mapone^2 \Maptwo \rag_c .
\label{H21}
\eeq
For the shear components $\gammais$ we have two new cross-correlations at
fourth order, $M_{31}$ (and its symmetric $M_{13}$) and $M_{22}$, and we 
define:
\beq
H_{31} = M_{31} - 3 \cM_{11} M_{20} - 3 \cM_{20} M_{11} + 3 \cM_{11} \cM_{20} ,
\;\; \mbox{whence} \;\; \lag\cH_{31}\rag = \lag\gamisone^3\gamistwo\rag_c = 
\lag\gamisone^3\gamistwo\rag - 3 \lag\gamisone^2\rag 
\lag\gamisone\gamistwo\rag ,
\label{H31}
\eeq
and:
\beq
H_{22} = M_{22} - \cM_{20} M_{02} - \cM_{02} M_{20} - 4 \cM_{11} M_{11} 
+ \cM_{20} \cM_{02} + 2  \cM_{11} \cM_{11} ,
\label{H22}
\eeq
whence:
\beq
\lag \cH_{22} \rag = \lag\gamisone^2\gamistwo^2\rag_c = 
\lag\gamisone^2\gamistwo^2\rag - \lag\gamisone^2\rag \lag\gamistwo^2\rag
- 2 \lag\gamisone\gamistwo\rag^2 .
\label{HH22}
\eeq
Again, one can easily check that the scatter of $H_{pq}$ is always smaller
than the scatter of $M_{pq}$ (see Appendix).
Finally, from these estimators $\cH_p$ and $\cH_{pq}$ we can evaluate the
parameters $S_p$, $t_p$ and the cross-correlation coefficients $r_{pq}$
introduced in eqs.(\ref{Sp}), (\ref{rpq}), through the estimators $\cS_p$, 
$\cT_p$ and $\cR_{pq}$:
\beq
\cS_p = \frac{\cH_p}{\cM_2^{p-1}} , \hspace{1cm} \cT_p = 
\frac{\cH_{p0}}{\cH_{0p}} \hspace{1cm} \mbox{and} \hspace{1cm}
\cR_{pq} = \frac{\cH_{p,q}}{\cH_{p+q,0}^{p/(p+q)} \cH_{0,p+q}^{q/(p+q)}} .
\label{cSp}
\eeq
We can note from numerical computations that the scatter of the moments
$\lag X_1^p X_2^q \rag$ increases very rapidly with the order $p+q$. Therefore,
we can neglect the contributions to the scatter of $\cS_p$ due to the 
denominator. Hence we write:
\beq
\lag \cS_p \rag \simeq S_p , \hspace{0.5cm} \sigma(\cS_p) \simeq 
\frac{\sigma(\cH_p)}{\lag X^2\rag^{p-1}} , \hspace{0.5cm}  
\lag \cT_p \rag \simeq t_p \hspace{0.5cm} \mbox{and} \hspace{0.5cm}
\lag \cR_{pq} \rag \simeq r_{pq} , 
\label{avcSp}
\eeq
\beq
\sigma^2(\cT_p) \simeq t_p^2 \left[ \frac{\sigma^2(\cH_{p0})}
{\lag X_1^p\rag_c^2} + \frac{\sigma^2(\cH_{0p})}{\lag X_2^p\rag_c^2} - 2 
\frac{\sigma^2(\cH_{p0};\cH_{0p})}{\lag X_1^p\rag_c \lag X_2^p\rag_c} \right] ,
\label{sigtp}
\eeq
and:
\beqa
\sigma^2(\cR_{pq}) & \simeq & r_{pq}^2 \Biggl \lbrace \frac{\sigma^2(\cH_{p,q})}
{\lag X_1^p X_2^q\rag_c^2} + \left(\frac{p}{p+q}\right)^2 
\frac{\sigma^2(\cH_{p+q,0})}{\lag X_1^{p+q}\rag_c^2} 
+ \left(\frac{q}{p+q}\right)^2  \frac{\sigma^2(\cH_{0,p+q})}
{\lag X_2^{p+q}\rag_c^2} - \frac{2p}{p+q} 
\frac{\sigma^2(\cH_{p,q};\cH_{p+q,0})}{\lag X_1^p X_2^q\rag_c 
\lag X_1^{p+q}\rag_c} \nonumber \\
& & - \frac{2q}{p+q} \frac{\sigma^2(\cH_{p,q};\cH_{0,p+q})}
{\lag X_1^p X_2^q\rag_c \lag X_2^{p+q}\rag_c} + \frac{2pq}{(p+q)^2} 
\frac{\sigma^2(\cH_{p+q,0};\cH_{0,p+q})}
{\lag X_1^{p+q}\rag_c \lag X_2^{p+q}\rag_c} \Biggl \rbrace .
\label{sigrpq}
\eeqa
Here we assumed that the scatter of $\cT_p$ and $\cR_{pq}$ is small so that it 
can be linearized in eqs.(\ref{cSp}). If this is not the case, that is
$\sigma^2(\cT_p) \ga 1$ or $\sigma^2(\cR_{pq}) \ga 1$, the error-bar obtained 
from eq.(\ref{sigtp}) or eq.(\ref{sigrpq}) is not correct (it significantly 
underestimates the scatter) but this means that the measure is dominated by 
the noise so that we do not need an accurate estimate of $\sigma^2$: we only 
wish to be aware that the noise is too large to allow a precise measure of 
$t_p$ or $r_{pq}$. In eqs.(\ref{sigtp})-(\ref{sigrpq}) we introduced
the cross-dispersions $\sigma^2(\cH_{p,q};\cH_{p',q'})$ defined by:
\beq
\sigma^2(\cH_{p,q};\cH_{p',q'}) = \lag \cH_{p,q} \cH_{p',q'} \rag 
- \lag \cH_{p,q} \rag \lag \cH_{p',q'} \rag .
\label{sigmacross}
\eeq
We give in appendix \ref{scatter} the expression of the various dispersions
$\sigma^2$ which we need in this article.

\section{Numerical Results}
\label{numres}

In this section we numerically evaluate the signal to noise ratios involved 
with the various estimators described in the previous section. We assume a 
specific model for the background cosmology, a specific correlation hierarchy 
for the matter distribution and we focus on the SNAP observational strategy 
for numerical calculations. However as mentioned before the basic formalism 
developed here remains completely general and specific details studied here 
only serve illustrational purposes.

\subsection{Cosmological Parameters}
\label{Cosmological Parameters}

For the background cosmology we consider a fiducial LCDM model with
$\Om=0.3$, $\Ol=0.7$, $H_0=70$ km/s/Mpc and $\sigma_8=0.88$. To check the
sensitivity of cross-correlations on the background dynamics of the universe
we shall also study the effect of small variations of these parameters 
onto weak-lensing observables and their cross-correlations.
We consider many-body correlations of the matter density field which
obey the stellar model (Valageas et al. 2004; Munshi et al. 2004). This is
actually identical to the minimal tree-model up to third-order moments
(Munshi et al. 2004). This is also coupled to the fit to the non-linear 
power-spectrum $P(k)$ of the dark matter density fluctuations 
given by Peacock \& Dodds (1996). Lowest order non-Gaussian signatures
will be at third order for $\Map$ and at fourth order for smoothed
shear components $\gammais$.

\subsection{Survey Parameters}
\label{Survey Parameters}

Hereafter, we adopt the characteristics of the SNAP mission as given in 
Refregier et al.(2004). More precisely, we consider the ``Wide'' survey where 
the redshift distribution of galaxies is given by:
\beq
n(z_s) \propto z_s^2 \; e^{-(z_s/z_0)^2} \;\;\; \mbox{and} \;\;\; 
z_0=1.13, \;\;\; z_{\rm max}=3 .
\label{nzSNAP}
\eeq
The variance in shear due to intrinsic ellipticities and measurement errors is 
$\sigma_*=\lag|\epsilon_*|^2\rag^{1/2}=0.31$. The survey covers an area 
$A=300$ deg$^2$ and the surface density of usable galaxies is 
$n_g=100$ arcmin$^{-2}$. Therefore, we take for the 
number $N$ of galaxies within a circular field of radius $\theta_s$:
\beq
N= n_g\pi\theta_s^2 \simeq 314 \left( \frac{n_g}{100 \mbox{arcmin}^{-2}} 
\right)  \left(\frac{\theta_s}{1\mbox{arcmin}}\right)^2,
\label{Nsurvey}
\eeq
and for the number $N_c$ of cells of radius $\theta_s$:
\beq
N_c= \frac{A}{(2\theta_s)^2} = 2.7 \times 10^5 \left(\frac{A}{300\mbox{deg}^2}
\right) \left(\frac{\theta_s}{1\mbox{arcmin}}\right)^{-2} .
\label{Ncsurvey}
\eeq
For the shear this number somewhat overestimates $N_c$ because of the 
sensitivity of $\gammais$ to long wavelengths, which would require 
the centres of different cells to be separated by more than $2\theta_s$ in 
order to be uncorrelated.

In order to extract some information from the redshift dependence of weak 
lensing effects, which could also be used to discriminate noise sources,
we also divide the ``Wide'' SNAP survey into two redshift bins. Then,
we shall study the cross correlations between both bins.
Thus, we consider the two subsamples which can be obtained from the
``Wide'' survey by dividing galaxies into two redshift bins: $z_s>z_*$
and $z_s<z_*$. We choose
$z_*=1.23$, which corresponds roughly to the separation provided by the SNAP
filters and which splits the ``Wide'' SNAP survey into two samples with the 
same number of galaxies (hence $n_g=50$ arcmin$^{-2}$). Note that one 
cannot use too many redshift bins at it decreases the number of source 
galaxies associated with each subsample (for the aperture mass we could still
obtain good results with three bins but we shall restrict ourselves to two
redshift bins in this paper). The redshift bins that we use are similar to 
those which are used by Refregier et al.(2003) using photometric redshifts, 
except that they have a sharp cutoff and non-overlapping source distributions. 
Note that using overlapping source distributions (over redshift) would 
increase the cross-correlations.

\subsection{Numerical evaluation of Estimators and their scatter} 

For all our computations we use the matter correlation hierarchy introduced 
and tested in a recent series of papers for the evaluation of statistics 
related to shear and $\Map$ statistics. These analytical models were tested 
extensively against numerical simulations and were found to provide 
satisfactory results. We have studied the cross correlations among the two 
redshift bins obtained from the ``Wide'' SNAP survey using both the aperture
mass $\Map$ and the smoothed shear components $\gammais$. In the following,
in weak-lensing observables of the form $H_{pq}$ the first index $p$ refers
to the low-$z$ subsample and the second index $q$ to the high-$z$ subsample.

\subsubsection{Aperture mass statistics $\Map$}
\label{Aperture-mass}

\paragraph{Second-order cumulants}
\label{Map-Second-order}

\begin{figure}
\protect\centerline{
\epsfysize = 2. truein
\epsfbox[22 516 591 705]
{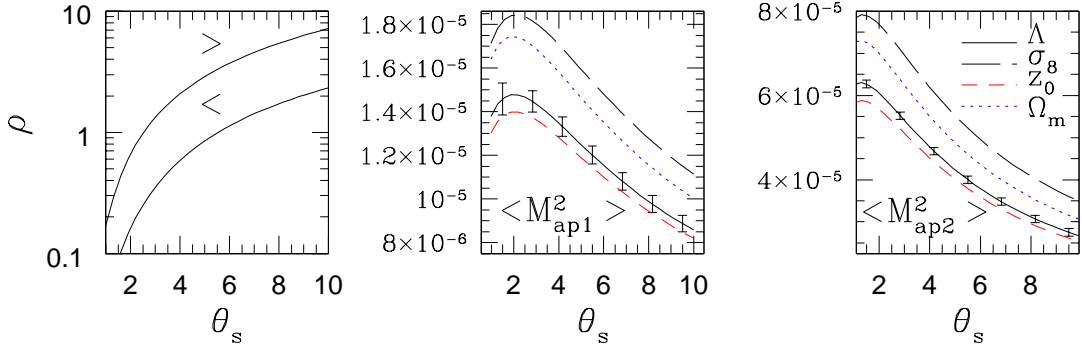}}
\caption{Left panel shows the quantity $\rho(\theta_s)$ which measures the
relative importance of galaxy intrinsic ellipticities, for the aperture mass
and two redshift bins, as a function of the angular scale $\theta_s$ 
(in arcmin). The variance $\lag\Map^2\rag$ for these two redshift bins is
displayed in middle (low-$z$ subsample) and right (high-$z$ subsample) panels.
Various line styles correspond to various cosmological or survey parameters.
The solid ``$\Lambda$'' line is our fiducial model described in sections
\ref{Cosmological Parameters}-\ref{Survey Parameters}. The dotted ``$\Om$'' 
curve shows the effect of a $10\%$ increase of $\Om$ (from $\Om=0.3$ up to
$\Om=0.33$), the dot-dashed ``$\sigma_8$'' line corresponds to a 
$10\%$ increase of $\sigma_8$ (from $\sigma_8=0.88$ up to $\sigma_8=0.97$)
while the dashed ``$z_0$'' curve represents a $10\%$ decrease of the
characteristic redshift $z_0$ of the survey (from $z_0=1.13$ down to 
$z_0=1.02$). These line styles will be used in the following plots.
The error bars show the one $\sigma$ dispersion around the mean value.}
\label{fig2Maprho}
\end{figure}

\begin{figure}
\protect\centerline{
\epsfysize = 2. truein
\epsfbox[22 433 591 705]
{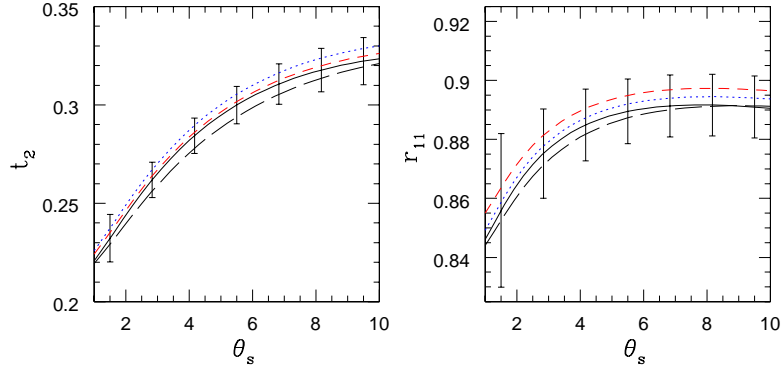}}
\caption{The quantities $t_2$ and $r_{11}$ (see text for description) for 
$\Map$ are plotted as a function of smoothing angle $\theta_s$ for the two 
redshift bins. Error bars correspond again to the one $\sigma$ dispersion 
around the mean value.}
\label{fig2Maptr}
\end{figure}

We first present in the left panel of Fig.\ref{fig2Maprho} the quantity 
$\rho(\theta_s)$ as a function of the angular smoothing radius $\theta_s$ 
for the two redshift bins obtained from the ``Wide'' SNAP survey. We can 
note that $\rho$ is of order unity which means that the galaxy intrinsic 
ellipticities yield a significant contribution to the signal which cannot 
be neglected (but weak lensing effects can still be extracted). The relative 
importance of intrinsic ellipticities decreases at large scales ($\rho$ 
increases) as the number $N$ of galaxies within the angular radius $\theta_s$ 
grows (this effect dominates over the decrease of the amplitude of weak 
lensing distortions, see eq.(\ref{rho})). The quantity $\rho$ is larger for
the high-redshift subsample because its variance $\lag\Map^2\rag_c$ is
larger (see eq.(\ref{rho})).

Next, we show in the middle and right panels of Fig.\ref{fig2Maprho} the 
variance $\lag\Map^2\rag$ of the aperture mass for each redshift subsample. 
The four curves correspond to our fiducial LCDM model and to a 
$10\%$ variation of $\Om$, $\sigma_8$ or $z_0$. We can see that the 
error-bars are quite small and would allow an accurate measure of the 
cosmological parameters. However, the signal is also sensitive to the 
redshift distribution of the sources which may be an important limiting 
factor (although the dependence is rather weak). As is well known, the 
amplitude of weak lensing distortions increases with the redshift of the 
sources as the line of sight is longer, hence the variance is larger for the 
high-$z$ subsample. On the other hand, it decreases at larger angular scales 
where the amplitude of density fluctuations is smaller. Note that the 
error-bars for the variance are quite small which justifies the approximation 
(\ref{avcSp}) for the scatter of $S_p$.

Then, we display in the left panel of Fig.\ref{fig2Maptr} the ratio 
$t_2$ between the variances $\lag\Map^2\rag$ of the weak lensing effects 
associated with each redshift bin, as defined in eq.(\ref{Sp}). In agreement 
with Fig.\ref{fig2Maprho} we have $t_2<1$. We can see that the
sensitivity to cosmological parameters is rather modest. The 
error-bars are of the same order as the variation with a $10\%$ change of 
cosmological parameters so that by combining several angular scales one 
can get a useful constraint on cosmology. In particular, note that $\Om$
and $\sigma_8$ affect the ratio $t_2$ in opposite directions while they
affect the variances $\lag\Map^2\rag$ in the same direction. Thus, 
dividing weak lensing surveys like the SNAP mission into several redshift 
bins provides additional information in addition to the overall amplitude 
of weak lensing effects averaged over all source redshifts.

Finally, we present in the right panel of Fig.\ref{fig2Maptr} the 
cross-correlation coefficient $r_{11}$ (defined in eq.(\ref{rpq})) 
between the two redshift bins, for 
our four models. We can see that the cross-correlation is quite strong 
($r_{11}\simeq 0.9$) with a small scatter. Indeed, the lines of sight 
associated with the two subsamples probe the same density fluctuations at 
low $z$, where they are largest because of the build-up of large-scale 
structures. Moreover, the cross-correlation coefficient $r_{11}$ only shows 
a very weak dependence on cosmology or the redshift distribution of sources: 
as the two subsamples are highly correlated most of the dependence on 
cosmology or source redshifts which appeared in Fig.\ref{fig2Maprho} cancels 
out between the numerator and denominator in eq.(\ref{rpq}). This also leads 
to reasonably small error-bars since the dispersions of the numerator and 
denominator do not merely add up in quadrature, see eq.(\ref{sigrpq}). 
The fact that the dependence on cosmology is very weak (and smaller than 
error-bars) means that the cross-correlation coefficient $r_{11}$ and the 
cross-product $\lag\Mapone\Maptwo\rag$ are not useful to constrain 
cosmological parameters. They do not add much information about cosmology 
to the one already included in the one-point statistics $\lag\Mapone^2\rag$ 
and $\lag\Maptwo^2\rag$. However, we can take advantage of this property to 
use $r_{11}$ to measure the observational noise. Indeed, since $r_{11}$ can 
be predicted up to a very good accuracy (e.g. from other statistics like the 
one-point variance within the same survey, or from other sources like the CMB) 
its value gives a simple estimate of the scatter of second-order cumulants 
as measured in a specific survey. It could also help track down systematics. 
As noticed in section~\ref{estimators}, eq.(\ref{sigrpq}) for the scatter of 
the cross-correlation coefficient $r_{11}$ assumed that the dispersion was 
small $\sigma \ll 1$. If this is not the case, that is the value $r_{11}$ 
obtained from a weak-lensing survey shows a large departure from its 
theoretical mean (e.g., one measures $r_{11} \la 0.5$), then eq.(\ref{sigrpq}) 
is an underestimate. Besides, the probability distribution of $r_{11}$ would 
be strongly non-Gaussian. Then, the estimate of the dispersion $\sigma$ of 
second-order cumulants one would derive from the measure of $r_{11}$ would 
be too large. Therefore, the break-up of eq.(\ref{sigrpq}) in this domain 
is not a real problem since it merely yields conservative estimates for the 
noise (besides in such cases the data is too noisy to provide accurate 
constraints on cosmology).

\paragraph{Third-order cumulants}
\label{Map-Third-order}

\begin{figure}
\protect\centerline{
\epsfysize = 2. truein
\epsfbox[22 433 591 705]
{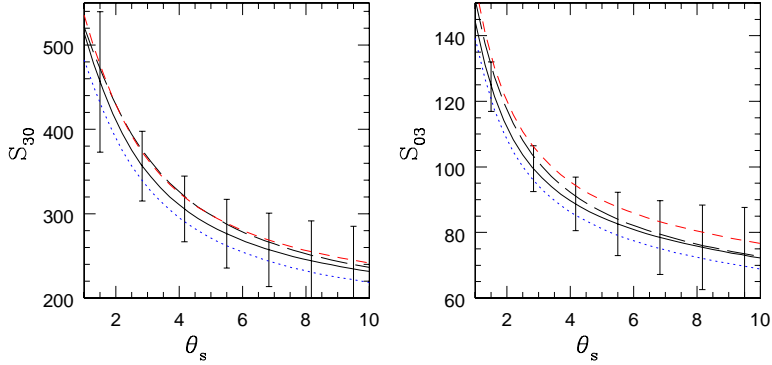}}
\caption{The skewness $S_3$ of the aperture mass is displayed as a function 
of the smoothing angle $\theta_s$, for the low-$z$ subsample (left panel)
and the high-$z$ subsample (right panel).}
\label{fig2MapS3}
\end{figure}

\begin{figure}
\protect\centerline{
\epsfysize = 2. truein
\epsfbox[22 516 591 705]
{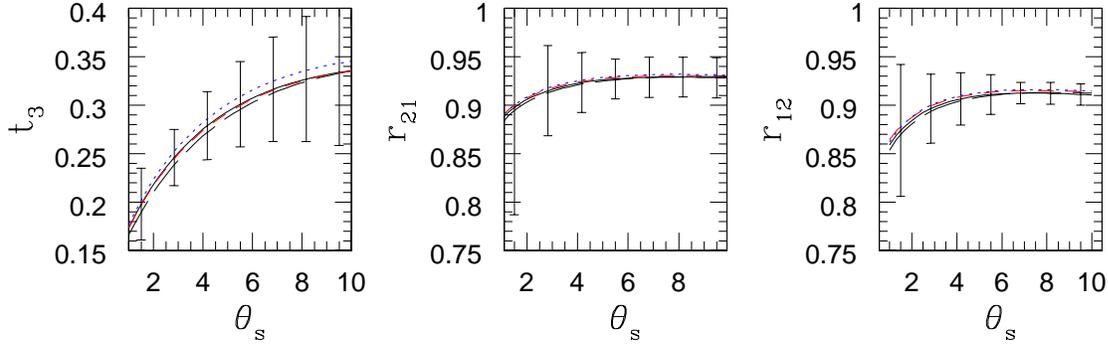}}
\caption{The ratios $t_3$, $r_{21}$ and $r_{12}$, for the $\Map$ statistic
and two redshift bins, are plotted in left, middle and right panel 
respectively. Error bars correspond again to one $\sigma$ scatter. Note the
very weak dependence on cosmology or source redshift.}
\label{fig2Mapt3r}
\end{figure}

The variance of weak lensing observables cannot simultaneously constrain 
several cosmological parameters. Thus, in order to break the $\Om-\sigma_8$
degeneracy it was proposed to consider the skewness $S_3$ 
(Bernardeau et al. 1997). More generally, taking into account higher-order
cumulants brings further information which can help constrain the underlying
cosmology. However, the noise increases very rapidly with the order of the
observed cumulants so that we shall restrict ourselves to the lowest-order
non-vanishing cumulants beyond second-order. For the aperture mass this
simply corresponds to third-order statistics like the skewness $S_3$ defined
in eq.(\ref{Sp}) (note that another possibility is to look at the whole 
probability distribution itself, see Valageas et al. 2004). We present in
Fig.\ref{fig2MapS3} the skewness $S_3$ of the aperture mass
as a function of the smoothing angle $\theta_s$ for the two redshift bins.
In agreement with previous works we find that the skewness increases for
lower source redshifts (because of the smaller length of the line of sight and
of the growth with time of the non-Gaussianities of the density field). Hence
we can check that $S_{30} > S_{03}$. We can see that although the scatter is
much larger than for the variance, one should obtain a clear detection of
non-Gaussianity in all cases. Moreover, by combining different angular scales
and redshift bins (e.g. through a Fisher matrix approach) one should be able
to constrain cosmological parameters at a $10\%$ level (note however the large
dependence on the source redshift, which is more important than for 
second-order cumulants).

We present in the left panel of Fig.\ref{fig2Mapt3r} the ratio $t_3$ between 
the third-order weak-lensing cumulants associated with the two redshift bins. 
In agreement with Fig.\ref{fig2MapS3} we obtain $t_3<1$. We note that
the dependence on cosmology is very small and much below error-bars.
Therefore, redshift binning is not very useful to constrain cosmological
parameters when it is applied to these third-order statistics. On the other
hand, the measure of $t_3$ could be used to evaluate the noise of 
third-order cumulants (or the skewness) since its value can be predicted with
a good accuracy (without requiring a high accuracy for cosmological 
parameters). This can be useful to assess the reliability of the data.

Next, we display in the middle and right panels of Fig.\ref{fig2Mapt3r} the 
cross-correlation coefficients $r_{21}$ and $r_{12}$ between both subsamples,
as defined in eq.(\ref{rpq}). We can check that the cross-correlation is
very strong ($r \sim 0.9$) which means that the weak lensing effects associated
with the two redshift bins are highly correlated. Of course, this is consistent
with the high value already obtained for the cross-correlation $r_{11}$
of second-order cumulants (Fig.\ref{fig2Maptr}). Note that we get 
$r_{21} > r_{12} > r_{11}$. Indeed, higher-order cumulants give more weight
to the common low-$z$ parts of the line of sights (because the parameters
$S_p$ of the 3-d density field increase in the non-linear regime). This
increases the correlation of higher-order weak lensing cumulants. Similarly,
the coefficient $r_{21}$ is slightly larger than $r_{12}$ as it gives more
weight to low-$z$ distortions. We can note that the scatter of the
cross-correlation coefficients $r_{21}$ and $r_{12}$ is somewhat smaller than
might be guessed from the dispersion of third-order cumulants 
(Fig.\ref{fig2MapS3}). This is again due to their high
correlation. As for the second-order cross-correlation $r_{11}$, the
dependence on cosmology or the source redshifts is very small (and much below
error-bars). This means that $r_{21}$ and $r_{12}$, as the cross-products
$\lag\Mapone^2\Maptwo\rag_c$ and $\lag\Mapone\Maptwo^2\rag_c$, do not bring
additional information to the one-point cumulants $S_3$ with regard to 
cosmology. However, even though they are mostly useless with respect
to the measure of cosmological parameters, they can be very useful to evaluate
the noise of third-order cumulants. Indeed, a simple measure of $r_{21}$ or
$r_{12}$ allows a good estimate of the noise of the skewness since their
theoretical value can be predicted with a good accuracy. Therefore, 
cross-correlations are a complementary tool to one-point statistics
and should prove useful.

\subsubsection{Smoothed shear components $\gammais$}
\label{shear}

We now describe our results for the shear components, following the outline
of our study of the aperture mass.

\paragraph{Second-order cumulants}
\label{shear-Second-order}

\begin{figure}
\protect\centerline{
\epsfysize = 2. truein
\epsfbox[22 500 591 705]
{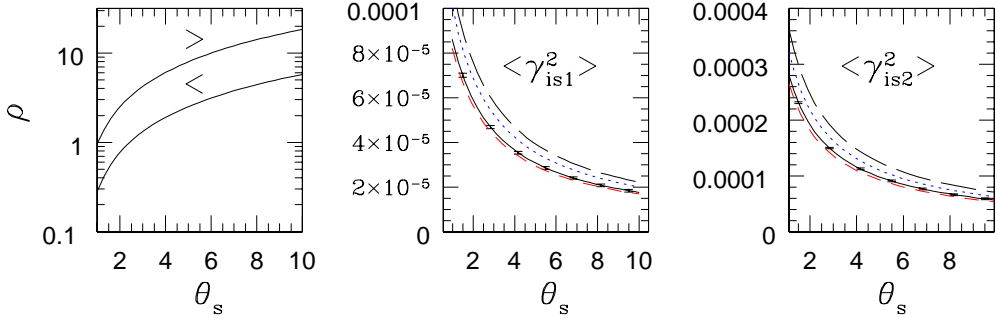}}
\caption{Same as Fig.\ref{fig2Maprho} but for shear components. Left 
panel shows the characteristic function $\rho$ associated with both
redshift bins. Middle and right panels show the variance $\lag\gammais^2\rag$ 
as a function of smoothing angle $\theta_s$ for lower and higher redshift bins 
respectively.}
\label{fig2shearrho}
\end{figure}

\begin{figure}
\protect\centerline{
\epsfysize = 2. truein
\epsfbox[22 430 591 705]
{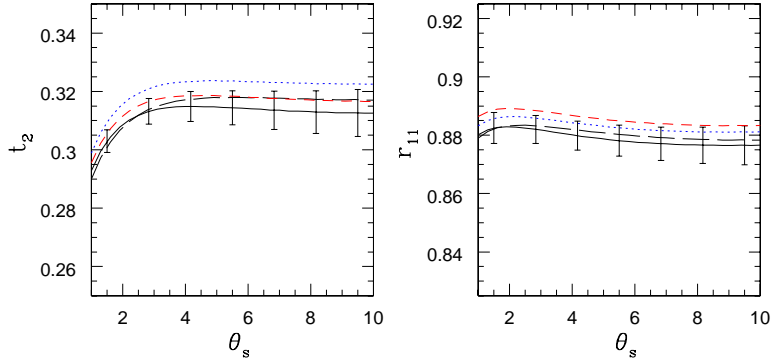}}
\caption{Same as Fig.\ref{fig2Maptr} but for shear components. As in case 
of $\Map$, $t_2$ and $r_{11}$ are almost independent of cosmological 
parameters.}
\label{fig2sheartr}
\end{figure}

We first show in the left panel of Fig.\ref{fig2shearrho} the quantity
$\rho$ which measures the importance of galaxy intrinsic ellipticities, while
we show the variance $\lag\gammais^2\rag$ in the middle panel (low-$z$ bin)
and the right panel (high-$z$ bin). As is well known, the variance of the
shear components is larger than the variance of the aperture mass because the
compensated filter of $\Map$ damps the contributions from long wavelengths.
This also leads to a larger value of $\rho$ for the shear components. Hence
the relative size of the error bars associated with second-order statistics
are smaller for the shear. The magnitude of these effects depends on the
slope of the matter density power-spectrum and they are larger at smaller
angular scales. Again, we can check that the error bars are very small and 
the variance associated with each bin should be accurately measured. This 
would yield a strong constraint on cosmology. 

Next, we present in the left panel of Fig.\ref{fig2sheartr} the ratio $t_2$
of both variances $\lag\gammais^2\rag$ associated with the two redshift
subsamples. Although the dependence on cosmology is rather weak, it could
be extracted from the data. Contrary to the aperture mass, $\Om$ and 
$\sigma_8$ affect the ratio $t_2$ in the same direction as each variance
$\lag\gammais^2\rag$ but with different powers. Therefore, redshift binning
still provides additional constraints on cosmology.
Finally, we display in the right panel of Fig.\ref{fig2sheartr} the 
cross-correlation coefficient $r_{11}$. As for $\Map$ it is rather high, which
translates a strong correlation between both weak-lensing signals, and it only
shows a very weak dependence on cosmology. However, since error bars are quite
small it might be possible to use $r_{11}$ to further constrain cosmological
parameters. On the other hand, this also means that it does not provide
a very robust tool to estimate the noise of the survey, unless the cosmological
parameters and the source redshifts are known up to a high accuracy.

\paragraph{Fourth-order cumulants}
\label{Shear-Fourth-order}

\begin{figure}
\protect\centerline{
\epsfysize = 2. truein
\epsfbox[22 430 591 705]
{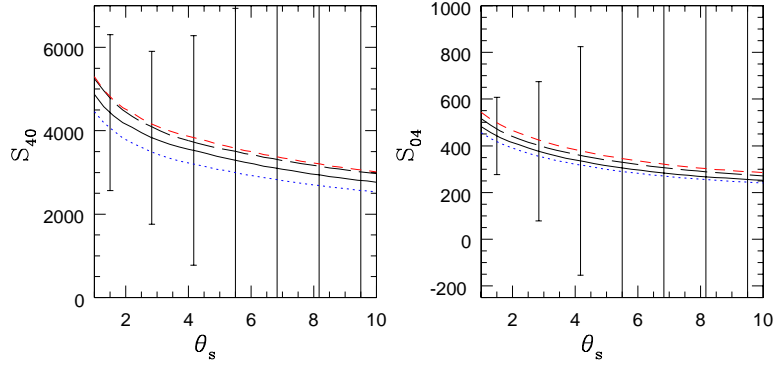}}
\caption{The kurtosis of smoothed shear components $S_4$ is plotted for two 
redshift bins as a function of smoothing angle $\theta_s$ (low-$z$ subsample:
left panel, high-$z$ subsample: right panel).}
\label{fig2shearS4}
\end{figure}

\begin{figure}
\protect\centerline{
\epsfysize = 2. truein
\epsfbox[22 500 591 705]
{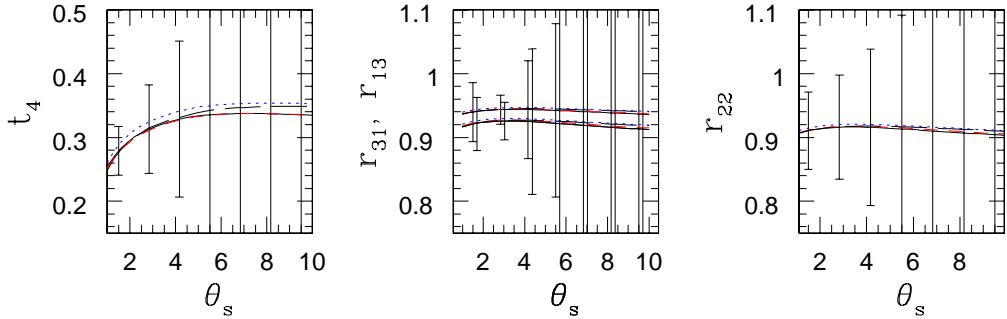}}
\caption{The left panel shows $t_4$ as a function of smoothing angle 
$\theta_s$. The middle and right panels correspond to $r_{31}, r_{13}$ and 
$r_{22}$ respectively. These quantities are related to the kurtosis of the
shear field and are dominated by the noise. We have reduced the error bars 
by a factor of four for plotting. }
\label{fig2sheart4r}
\end{figure}

As for $\Map$, we now study the lowest-order cumulants which probe the
deviations from Gaussianity. Since the shear components are even random
variables we need to go up to fourth order. Since the noise increases very
fast with the order this means that non-Gaussianities are more difficult to
detect from the shear components than from $\Map$ (where third-order cumulants
do not vanish). Thus, we display in Fig.\ref{fig2shearS4} the kurtosis $S_4$
of the smoothed shear components for both redshift subsamples. As for the
skewness of $\Map$, the kurtosis increases for lower source redshifts, and
we obtain $S_{40} > S_{04}$. As expected, we can check that the error bars
are much larger than for the skewness of $\Map$. In particular, at large
angular scales (beyond $6'$) it will be difficult to obtain a clear detection
of non-Gaussianity. At lower angular scales one should be able to obtain
a meaningful measure of $S_4$. On the other hand, one should note that the
theoretical error bars for $S_4$ are actually rather large in this transition
regime towards non-linear gravitational structures, so that one cannot expect
to obtain strong constraints on cosmology from $S_4$. Rather, it is probably
safer to use $S_4$ to check the broad consistency of the gravitational 
clustering process.

Next, we show in the left panel of Fig.\ref{fig2sheart4r} the ratio $t_4$
between the fourth-order cumulants associated with both redshift subsamples.
As for $\Map$, we get $t_4<1$ and the dependence on cosmology is very small.
Again, this means that redshift binning does not provide much additional
information for fourth-order statistics. On the other hand, a measure of
$t_4$ could provide a convenient estimate of the noise of the survey with
regard to these fourth-order cumulants.

Finally, we present in the middle and right panels of Fig.\ref{fig2sheart4r}
the cross-correlation coefficients $r_{31}, r_{13}$ and $r_{22}$. We obtain
$r_{31}>r_{13}>r_{22}$. Again, the cross-correlation is large and the 
dependence on cosmology is very weak and much smaller than error bars.
Therefore, these cross-correlations are mostly useful to estimate the noise
of the survey rather than the cosmological parameters. We must note that at
large angular scales the formula (\ref{sigrpq}) for $\sigma^2(\cR_{pq})$
yields a negative number (in which case with plot $\sigma=|\sigma^2|^{1/2}$.
This translates the fact that error bars are so large that errors cannot be
linearized in the expression (\ref{cSp}) of $\cR_{pq}$. However, in these cases
we have $\sigma \ga 1$ so that an accurate estimate of the scatter is not 
needed: the signal is completely dominated by the noise and useless for
practical purposes.

\section{Discussion and Future Prospects}


In order to derive precise results from weak lensing surveys one must take into
account various sources of noise like the intrinsic ellipticity distribution
of galaxies and the cosmic variance. Extending a previous study 
(Valageas, Munshi \& Barber 2004) we have introduced these realistic sources 
of noise in various estimates of cross-correlation statistics between several
redshift subsamples. Such studies should help discriminating weak lensing
effects from noise and measuring cosmological parameters.


Thus, extending an earlier work (Valageas, Munshi \& Barber 2004) where
we focused on one-point statistics, we have taken advantage of the high 
projected density of galaxies $n_g$ in the future SNAP survey to divide the
source population into two redshift bins. Then, we have computed
cross-correlations between these subsamples. Note that our formalism is kept
completely general and can be applied to different surveys with or without any 
overlap in source redshift. We have used two statistics, the aperture mass
$\Map$ and the shear components $\gammais$, to quantify the correlations among 
redshift bins as a function of the smoothing angle $\theta_s$. The underlying
model that we have used for computing the cumulant correlators is the same as 
that of Valageas, Munshi \& Barber (2004). This has been tested against 
numerical simulations for a wide range of smoothing angles and source 
redshifts and was found to provide good results.

We have considered both second-order cumulants (cross-correlation $r_{11}$) and
the lowest-order cumulants which can measure deviations from Gaussianity
(third or fourth order for $\Map$ or $\gammais$). High order cumulants can
remove the well-known degeneracy between $\Om$ and $\sigma_8$ but the relative
importance of the noise increases rapidly with order. This makes the aperture
mass a better tool than the shear components which require fourth-order
statistics. However, note that weak lensing statistics are also sensitive 
to source redshifts which might prove a significant limitation for any 
weak lensing survey. We noticed that for a survey such as SNAP one could
still obtain interesting information from the shear components after dividing
the sample into two redshift bins, while for the aperture mass one could go
up to three subsamples. Increasing further the number of subsamples increases
the noise of third-order or higher order cumulants and is mostly useless
(except for the variance where the error bars are small). We have found that
cross-correlations between subsamples do not bring additional information 
with regard to the constraints on cosmology to those already provided by 
one-point statistics (except marginally for the second-order moments of the
shear). Indeed, different source populations selected by their redshift
(in the zero angular separation case) still probe the same density 
fluctuations at low $z$ (where they are largest). This implies a high value
for the various cross-correlation coefficients which are almost independent 
of cosmological parameters. These properties hold even better for higher 
orders statistics. However, this feature can be used to estimate the
noise of the survey. Therefore, one-point and two-point statistics
provide complementary tools: the former can constrain cosmology and check
the gravitational clustering process, while the latter can yield an estimate
of the noise of the survey.


Recent studies have underlined the importance of joint estimation of 
cosmological parameters such as galaxy-shear cross correlation, shear-shear 
correlations along with galaxy-galaxy angular correlations. Deep multi-colour 
galaxy surveys with photometric redshifts will provide an unique opportunity 
to study such large numbers of two-point correlation observables to pin-point 
the background dynamics of the universe as well as non-Gaussianity and bias 
associated with cosmological density distribution of underlying mass as well 
as galaxy populations. The statistical machinery developed here will be 
extremely useful in this direction.

Although the cross correlations and the error terms are defined in this work 
with a specific filter in mind and we only concentrate on two different 
statistics the framework developed here has been kept completely general and 
the formalism can be used to compute the signal to noise related to quantities 
such as foreground-background galaxy populations (Moessner \& Jain 1998). 
Fluctuations in size distribution of source galaxies can also be used to study 
the weak lensing convergence directly without having to go through the map 
making procedure, from shear maps to convergence maps (see e.g. Jain, 2002). 
Higher order cross-correlation statistics or cumulant correlators we develop 
here can be useful in probing correlations of such measures of weak lensing 
convergence with cosmic shear measurements using shapes of background galaxies 
as observations improve and surveys start to cover a larger part of the sky 
and probe deeper. Not only these statistics can provide valuable consistency 
checks among various measures from different surveys, they will also be a 
useful supplement in constraining cosmological parameters when used along with 
their one-point counterparts.


Recently Dalal et al. (2002) have investigated the correlation between
the smoothed convergence and the value along the line of sight to individual 
supernovae. However these calculations were done by ignoring non-Gaussian 
corrections. The formalism developed here takes into account all non-linear 
corrections from higher order contributions. In our formalism we have shown
that such calculations can be treated as a special case of the generic
expression derived here in a more general context. These results can not only 
be used to analyze cross-correlations among various redshift surveys but 
also to probe supernova weak-lensing cross-correlation. A more detailed 
analysis will be presented elsewhere.


Although current studies of weak lensing effects have focused on the
distortion of background galaxy ellipticities, it has been pointed out recently
that such studies can be augmented in the near future by lensing studies of 
unresolved sources and also of the spatial fluctuations of their integrated 
diffuse emission. In particular one could study the diffuse background generated
at far-infrared wavelengths by dusty star bursts, first stars and galaxies in 
near-infrared and also the 21cm emission from neutral gas which forms 
the intergalactic medium prior to reionization. These studies promise to 
smoothly interpolate from the weak-lensing studies of the cosmic 
microwave background at very high redshift downto weak lensing studies using 
galaxy shapes at low redshifts. Cross-correlation studies of one type of 
surveys with another hold the prospect of mapping the dark matter distribution 
of the universe at medium redshift. We plan to present results of such analysis 
in a separate paper.


In almost all weak lensing studies one uses the Born-approximation.
Its validity is checked in numerous numerical studies
of weak lensing at small angular scales. In addition we 
have ignored contributions from source clustering and lens coupling.
Some of these issues have been studied in Bernardeau (1997),
Bernardeau et al (1997) and Schneider et al. (1998). It was 
found that at large smoothing angular scales these 
contributions are negligible. In the highly nonlinear regime
an accurate picture of galaxy bias is required to deal with
these issues but it is still lacking at the moment.


The correlations of galaxy intrinsic ellipticities might yield
additional complexities in dealing with shear correlations and elaborate
schemes have been developed (Heymans \& Heavens 2003, Crittenden et al. 2002). 
However the extent to which such correlations will affect weak lensing 
surveys remains somewhat uncertain. It is generally believed that such
effects will play a less important role as we increase the survey depth
and we can reduce their role through acquisition of photometric redshift 
(Heymans \& Heavens 2003). 


It is almost always assumed in all weak lensing studies that the galaxy
intrinsic ellipticity distribution is Gaussian although it might not be the 
case. Any signature of such non-Gaussianity if found by observational teams 
will have to be included into our analytical calculations. 
However this can be performed in a straightforward way in our formalism.
Clearly, if such intrinsic non-Gaussianities are too large they might even
dominate the signal and complicate inferring the observational data.


It became clear that future weak lensing surveys will mostly be 
limited by sky coverage which determines the scatter due to the finite
size of the catalog. The intrinsic ellipticities will only dominate 
at small angular scales and will not preclude a meaningful estimation
even with small sky coverage. In our analysis we have used the plane 
parallel approximation to compute the higher order non-Gaussianities. 
However as the survey size increases
in the future it will be important to translate these studies using all-sky
calculations. A detailed analysis will be presented elsewhere.

\section*{acknowledgments}

DM was supported by PPARC of grant
RG28936. It is a pleasure for DM to acknowledge many fruitful
discussions with members of Cambridge Leverhulme Quantitative
Cosmology Group. We thank Andrew Barber for previous collaborations
which helped us to initiate the present study.

\appendix

\section{Scatter of low-order estimators $M_{pq}$ and $H_{pq}$}
\label{scatter}

We give in this appendix the expression of the scatter $\sigma^2$ of the
various estimators $M_{pq}$ and $H_{pq}$ introduced in 
section~\ref{estimators} which we need to derive the numerical results
presented in section~\ref{numres}. Note that although in our numerical 
calculations we take the smoothing windows $\theta_s$ of different surveys or
subsamples to be the same when we compute cross-correlations, this needs not 
to be the case for the following equations to be valid.

\subsection{Aperture Mass}

We first consider the aperture mass statistics, where we restrict ourselves
to cumulants of order two and three. The dispersions $\sigma^2$ of first and
second-order estimators are (with $\lag M_1\rag=0$):
\beq
\sigma^2(M_1) = \lag\Map^2\rag_c \left[1+{1\over\rho}\right] , \hspace{0.9cm} 
\sigma^2(M_2) = \lag\Map^4\rag_c + \lag\Map^2\rag_c^2
2 \left[1+{1\over\rho}\right]^2 ,
\label{sigM2}
\eeq
and:
\beq
\sigma^2(M_{11}) = \lag\Mapone^2\Maptwo^2\rag_c + \lag\Mapone\Maptwo\rag_c^2 
+ \lag\Mapone^2\rag_c \lag\Maptwo^2\rag_c \left[1+{1\over\rho_1}\right] 
\left[1+{1\over\rho_2}\right] .
\label{sigM11}
\eeq
Note that in the case $\rho_1 = \rho_2$ the cross-correlation error
$\sigma^2(M_{11})$ does not become identical to the error $\sigma^2(M_2)$
associated with one survey for the estimator $M_2$. This simply reflects
our assumption that the intrinsic ellipticities of different galaxies are
not correlated and that different surveys or subsamples do not share
any galaxy. Of course, it is also possible to handle the case where a 
certain fraction of the source population is shared by both surveys but 
we shall not consider this case here. As seen in eqs.(\ref{sigrpq}), 
(\ref{sigmacross}), we also need the dispersions of cross-products:
\beq
\sigma^2(M_{11};M_{20}) = \lag\Mapone^3\Maptwo\rag_c + \lag\Mapone^2\rag_c
\lag\Mapone\Maptwo\rag_c 2 \left[1+{1\over\rho_1}\right] , \hspace{0.8cm}
\sigma^2(M_{20};M_{02}) = \lag\Mapone^2\Maptwo^2\rag_c 
+ 2 \lag\Mapone\Maptwo\rag_c^2 .
\label{sigM11M20}
\eeq
Note that the galaxy intrinsic ellipticities do not contribute to the 
dispersion of the product $(M_{20} M_{02})$ because we assumed that the 
two subsamples have no common galaxies. Next, the scatters of third-order
estimators are:
\beq
\sigma^2(M_3) = \lag\Map^6\rag_c + \lag\Map^4\rag_c\lag\Map^2\rag_c 
\left[15+{9\over\rho}\right] + 9 \lag\Map^3\rag_c^2 + \lag\Map^2\rag^3 
\left[15+{27\over\rho}+{18\over\rho^2}+{6\over\rho^3}\right] , 
\label{sigM3}
\eeq
and:
\beqa
\sigma^2(M_{21}) & = & \lag\Mapone^4\Maptwo^2\rag_c + \lag\Mapone^4\rag_c 
\lag\Maptwo^2\rag_c \left[1+{1\over\rho_2}\right] 
+ 8 \lag\Mapone^3\Maptwo\rag_c \lag\Mapone\Maptwo\rag_c + 
\lag\Mapone^2\Maptwo^2\rag_c \lag\Mapone^2\rag_c \left[6+{4\over\rho_1}\right]
\nonumber \\ & & + 4 \lag\Mapone^3\rag_c \lag\Mapone\Maptwo^2\rag_c
+ 5 \lag\Mapone^2\Maptwo\rag_c^2 + \lag\Mapone^2\rag_c^2 \lag\Maptwo^2\rag_c
\left[3+{4\over\rho_1}+{3\over\rho_2}+{2\over\rho_1^2}+{4\over\rho_1\rho_2} 
+{2\over\rho_1^2\rho_2}\right] \nonumber \\ & & + \lag\Mapone\Maptwo\rag_c^2 
\lag\Mapone^2\rag_c \left[12+{8\over\rho_1}\right] .
\label{sigM21}
\eeqa
For cross-products we obtain:
\beqa
\sigma^2(M_{21};M_{30}) & = & \lag\Mapone^5\Maptwo\rag_c + 5 
\lag\Mapone^4\rag_c \lag\Mapone\Maptwo\rag_c + \lag\Mapone^3\Maptwo\rag_c
\lag\Mapone^2\rag_c \left[10+{6\over\rho_1}\right] + 9 
\lag\Mapone^2\Maptwo\rag_c \lag\Mapone^3\rag_c \nonumber \\
& & + \lag\Mapone\Maptwo\rag_c \lag\Mapone^2\rag_c^2 
\left[15+{18\over\rho_1}+{6\over\rho_1^2}\right] ,
\label{sigM21M30}
\eeqa
\beqa
\lefteqn{ \sigma^2(M_{21};M_{03}) = \lag\Mapone^2\Maptwo^4\rag_c 
+ \lag\Maptwo^4\rag_c \lag\Mapone^2\rag_c + 8 \lag\Mapone\Maptwo^3\rag_c
\lag\Mapone\Maptwo\rag_c + \lag\Mapone^2\Maptwo^2\rag_c \lag\Maptwo^2\rag_c
\left[6+{3\over\rho_2}\right] } \nonumber \\ & & 
+ 3 \lag\Mapone^2\Maptwo\rag_c \lag\Maptwo^3\rag_c + 6 
\lag\Mapone\Maptwo^2\rag_c^2 + \lag\Mapone^2\rag_c \lag\Maptwo^2\rag_c^2 
\left[3+{3\over\rho_2}\right] + \lag\Mapone\Maptwo\rag_c^2 \lag\Maptwo^2\rag_c 
\left[12+{6\over\rho_2}\right] ,
\label{sigM21M03}
\eeqa
and:
\beqa
\sigma^2(M_{30};M_{03}) & = & \lag\Mapone^3\Maptwo^3\rag_c + 3 
\lag\Mapone^3\Maptwo\rag_c \lag\Maptwo^2\rag_c + 3 \lag\Mapone\Maptwo^3\rag_c
\lag\Mapone^2\rag_c + 9 \lag\Mapone^2\Maptwo^2\rag_c \lag\Mapone\Maptwo\rag_c
\nonumber \\ & & + 9 \lag\Mapone^2\Maptwo\rag_c \lag\Mapone\Maptwo^2\rag_c
+ 9 \lag\Mapone^2\rag_c \lag\Maptwo^2\rag_c 
\lag\Mapone\Maptwo\rag_c + 6 \lag\Mapone\Maptwo\rag_c^3 .
\label{sigM30M03}
\eeqa
As recalled in section~\ref{estimators}, following Valageas et al. (2004),
it is better to use the cumulant estimators $H_{pq}$ rather than the moment
estimators $M_{pq}$, as they show a smaller scatter. Their dispersion can be
expressed in terms of the scatter of the estimators $M_{pq}$ as:
\beq
\sigma^2(H_3) = \sigma^2(M_3) - 6 \lag\Map^4\rag_c \lag\Map^2\rag_c 
- 9 \lag\Map^2\rag_c^3 \left[1+{1\over\rho}\right] ,
\label{sigH3}
\eeq
\beqa
\sigma^2(H_{21}) & = & \sigma^2(M_{21}) - 4 \lag\Mapone^3\Maptwo\rag_c 
\lag\Mapone\Maptwo\rag_c - 2 \lag\Mapone^2\Maptwo^2\rag_c \lag\Mapone^2\rag_c
- \lag\Mapone^2\rag_c^2 \lag\Maptwo^2\rag_c \left[1+{1\over\rho_2}\right]
\nonumber \\ & & - \lag\Mapone\Maptwo\rag_c^2 \lag\Mapone^2\rag_c 
\left[8+{4\over\rho_1}\right] ,
\label{sigH21}
\eeqa
\beq
\sigma^2(H_{21};H_{30}) = \sigma^2(M_{21};M_{30}) - 2 \lag\Mapone^4\rag_c 
\lag\Mapone\Maptwo\rag_c - 4 \lag\Mapone^3\Maptwo\rag_c \lag\Mapone^2\rag_c 
- \lag\Mapone\Maptwo\rag_c \lag\Mapone^2\rag_c^2 \left[9+{6\over\rho_1}\right] ,
\label{sigH21H30}
\eeq
\beqa
\sigma^2(H_{21};H_{03}) & = &  \sigma^2(M_{21};M_{03}) - \lag\Maptwo^4\rag_c 
\lag\Mapone^2\rag_c - 2 \lag\Mapone\Maptwo^3\rag_c \lag\Mapone\Maptwo\rag_c 
- 3 \lag\Mapone^2\Maptwo^2\rag_c \lag\Maptwo^2\rag_c \nonumber \\
& & - \lag\Mapone^2\rag_c \lag\Maptwo^2\rag_c^2 \left[3+{3\over\rho_2}\right] 
- 6 \lag\Mapone\Maptwo\rag_c^2 \lag\Maptwo^2\rag_c ,
\label{sigH21H03}
\eeqa
and:
\beq
\sigma^2(H_{30};H_{03}) = \sigma^2(M_{30};M_{03}) - 3 
\lag\Mapone^3\Maptwo\rag_c \lag\Maptwo^2\rag_c - 3 \lag\Mapone\Maptwo^3\rag_c
\lag\Mapone^2\rag_c - 9 \lag\Mapone^2\rag_c \lag\Maptwo^2\rag_c 
\lag\Mapone\Maptwo\rag_c .
\label{sigH30H03}
\eeq
We can check in eqs.(\ref{sigH3})-(\ref{sigH30H03}) that the scatter of the
estimators $H_{pq}$ is smaller than the dispersion of the estimators $M_{pq}$.
Besides, one can note that the galaxy intrinsic ellipticity dispersion only 
enters the expressions of the scatter of the estimators $H_{pq}$ through
the combination $(1+1/\rho)$.

\subsection{Shear components}

The dispersions of second-order cumulants are given by the same
expressions (\ref{sigM2})-(\ref{sigM11M20}) as the one obtained for the
aperture mass. On the other hand, since all odd-order moments of 
shear components vanish we need to go up to fourth-order moments in order 
to obtain the first measure of deviations from Gaussianity. Thus, we obtain
for fourth-order statistics:
\beqa
\sigma^2(M_4) & = & \lag\gammais^8\rag_c + \lag\gammais^6\rag_c 
\lag\gammais^2\rag_c \left[ 28+\frac{16}{\rho} \right] 
+ 34 \lag\gammais^4\rag_c^2 + \lag\gammais^4\rag_c \lag\gammais^2\rag_c^2 
\left[ 204 + \frac{240}{\rho} + \frac{72}{\rho^2} \right] \nonumber \\ 
&& + \lag\gammais^2\rag_c^4 \left[ 96 + \frac{240}{\rho} 
+ \frac{216}{\rho^2} + \frac{96}{\rho^3} + \frac{24}{\rho^4} \right] ,
\label{sigM4}
\eeqa
\beqa
\sigma^2(M_{31}) & = & \lag\gamisone^6\gamistwo^2\rag_c +  
\lag\gamisone^6\rag_c \lag\gamistwo^2\rag_c \left[1+{1\over\rho_2}\right] 
+ 12 \lag\gamisone^5\gamistwo\rag_c  \lag\gamisone\gamistwo\rag_c
+ \lag\gamisone^4\gamistwo^2\rag_c \lag\gamisone^2\rag_c 
\left[15+{9\over\rho_1}\right] \nonumber \\ 
&& + 15 \lag\gamisone^4\rag_c \lag\gamisone^2\gamistwo^2\rag_c + 19 
\lag\gamisone^3\gamistwo\rag_c^2 + \lag\gamisone^4\rag_c \lag\gamisone^2\rag_c 
\lag\gamistwo^2\rag_c \left[15+{9\over\rho_1}+{15\over\rho_2}+
{9\over\rho_1\rho_2}\right] + 30 \lag\gamisone^4\rag_c 
\lag\gamisone\gamistwo\rag_c^2 \nonumber \\
&& + \lag\gamisone^3\gamistwo\rag_c \lag\gamisone\gamistwo\rag_c 
\lag\gamisone^2\rag_c \left[114+{72\over\rho_1}\right] + 
\lag\gamisone^2\gamistwo^2\rag_c \lag\gamisone^2\rag_c^2
\left[45+{54\over\rho_1}+{18\over\rho_1^2}\right] \nonumber \\
&& + \lag\gamisone^2\rag_c^3 \lag\gamistwo^2\rag_c \left[15+{27\over\rho_1}
+{15\over\rho_2}+{18\over\rho_1^2}+{27\over\rho_1\rho_2}
+{6\over\rho_1^3}+{18\over\rho_1^2\rho_2}+{6\over\rho_1^3\rho_2}\right] 
+ \lag\gamisone\gamistwo\rag_c^2 \lag\gamisone^2\rag_c^2 \left[81
+{108\over\rho_1}+{36\over\rho_1^2}\right] ,
\label{sigM31}
\eeqa
and:
\beqa
\lefteqn{ \sigma^2(M_{22}) = \lag\gamisone^4\gamistwo^4\rag_c + 16
\lag\gamisone^3\gamistwo^3\rag_c \lag\gamisone\gamistwo\rag_c + 
\lag\gamisone^2\gamistwo^4\rag_c \lag\gamisone^2\rag_c \left[6+{4\over\rho_1}
\right] + \lag\gamisone^4\gamistwo^2\rag_c \lag\gamistwo^2\rag_c \left[6+ 
{4\over\rho_2}\right] + \lag\gamisone^4\rag_c \lag\gamistwo^4\rag_c } 
\nonumber \\ 
&& + 16 \lag\gamisone\gamistwo^3\rag_c \lag\gamistwo\gamisone^3\rag_c
+ 17 \lag\gamisone^2\gamistwo^2\rag_c^2 + 68 \lag\gamisone^2\gamistwo^2\rag_c
\lag\gamisone\gamistwo\rag_c^2 + \lag\gamisone^4\rag_c \lag\gamistwo^2\rag_c^2
\left[3+{4\over\rho_2}+{2\over\rho_2^2}\right] \nonumber \\
&& + \lag\gamistwo^4\rag_c \lag\gamisone^2\rag_c^2 \left[3+{4\over\rho_1}
+{2\over\rho_1^2}\right] + \lag\gamisone^3\gamistwo\rag_c 
\lag\gamisone\gamistwo\rag_c \lag\gamistwo^2\rag_c \left[48+{32\over\rho_2}
\right] + \lag\gamisone\gamistwo^3\rag_c \lag\gamisone\gamistwo\rag_c
\lag\gamisone^2\rag_c \left[48+{32\over\rho_1}\right] \nonumber \\
&& + \lag\gamisone^2\gamistwo^2\rag_c \lag\gamisone^2\rag_c 
\lag\gamistwo^2\rag_c \left[34+{24\over\rho_1}+{24\over\rho_2}
+{16\over\rho_1\rho_2}\right] + 20 \lag\gamisone\gamistwo\rag_c^4 
+ \lag\gamisone^2\rag_c \lag\gamistwo^2\rag_c \lag\gamisone\gamistwo\rag_c^2 
\left[68+{48\over\rho_1}+{48\over\rho_2}+{32\over\rho_1\rho_2}\right] 
\nonumber \\ && + \lag\gamisone^2\rag_c^2 \lag\gamistwo^2\rag_c^2 
\left[8+{12\over\rho_2}+{12\over\rho_1}+{6\over\rho_1^2}+{6\over\rho_2^2}
+{16\over\rho_1\rho_2}+{8\over\rho_1^2\rho_2}+{8\over\rho_2^2\rho_1}
+{4\over\rho_1^2\rho_2^2}\right].
\label{sigM22}
\eeqa
Next, for cross-products we obtain:
\beqa
\sigma^2(M_{31};M_{40}) & = & \lag\gamisone^7\gamistwo\rag_c + 7 
\lag\gamisone^6\rag_c \lag\gamisone\gamistwo\rag_c + 
\lag\gamisone^5\gamistwo\rag_c \lag\gamisone^2\rag_c 
\left[21+{12\over\rho_1}\right] + 34 \lag\gamisone^3\gamistwo\rag_c 
\lag\gamisone^4\rag_c \nonumber \\
& & + \lag\gamisone^3\gamistwo\rag_c \lag\gamisone^2\rag_c^2
\left[102+{120\over\rho_1}+{36\over\rho_1^2}\right] + \lag\gamisone^4\rag_c 
\lag\gamisone\gamistwo\rag_c \lag\gamisone^2\rag_c 
\left[102+{60\over\rho_1}\right] \nonumber \\ 
& & + \lag\gamisone\gamistwo\rag_c \lag\gamisone^2\rag_c^3 \left[96
+{180\over\rho_1}+{108\over\rho_1^2}+{24\over\rho_1^3}\right] ,
\label{sigM31M40}
\eeqa
\beqa
\lefteqn{ \sigma^2(M_{31};M_{04}) = \lag\gamisone^3\gamistwo^5\rag_c + 3 
\lag\gamisone\gamistwo^5\rag_c \lag\gamisone^2\rag_c + 
\lag\gamisone^3\gamistwo^3\rag_c \lag\gamistwo^2\rag_c \left[10+
{4\over\rho_2}\right] + 15 \lag\gamisone^2\gamistwo^4\rag_c 
\lag\gamisone\gamistwo\rag_c } \nonumber \\ 
&& + 4 \lag\gamisone^3\gamistwo\rag_c \lag\gamistwo^4\rag_c 
+ 30 \lag\gamisone^2\gamistwo^2\rag_c \lag\gamisone\gamistwo^3\rag_c 
+ \lag\gamisone^3\gamistwo\rag_c \lag\gamistwo^2\rag_c^2 
\left[12+{12\over\rho_2}\right] \nonumber \\
&& + 12 \lag\gamistwo^4\rag_c \lag\gamisone\gamistwo\rag_c 
\lag\gamisone^2\rag_c + \lag\gamisone\gamistwo^3\rag_c \lag\gamisone^2\rag_c 
\lag\gamistwo^2\rag_c \left[30+{12\over\rho_2}\right] + 60 
\lag\gamisone\gamistwo^3\rag_c \lag\gamisone\gamistwo\rag_c^2 \nonumber \\
&& + \lag\gamisone^2\gamistwo^2\rag_c \lag\gamisone\gamistwo\rag_c 
\lag\gamistwo^2\rag_c \left[90+{36\over\rho_2}\right] 
+ \lag\gamisone\gamistwo\rag_c \lag\gamisone^2\rag_c 
\lag\gamistwo^2\rag_c^2 \left[36+{36\over\rho_2}\right] + 
\lag\gamisone\gamistwo\rag_c^3 \lag\gamistwo^2\rag_c \left[60+{24\over\rho_2}
\right] ,
\label{sigM31M04}
\eeqa
\beqa
\sigma^2(M_{40};M_{04}) & = & \lag\gamisone^4\gamistwo^4\rag_c + 6
\lag\gamisone^2\gamistwo^4\rag_c \lag\gamisone^2\rag_c + 6 
\lag\gamisone^4\gamistwo^2\rag_c \lag\gamistwo^2\rag_c + 16 
\lag\gamisone^3\gamistwo^3\rag_c \lag\gamisone\gamistwo\rag_c + 16
\lag\gamisone^3\gamistwo\rag_c \lag\gamisone\gamistwo^3\rag_c \nonumber \\
&& + 18 \lag\gamisone^2\gamistwo^2\rag_c^2 + 48 \lag\gamisone\gamistwo^3\rag_c
\lag\gamisone^2\rag_c \lag\gamisone\gamistwo\rag_c + 48 
\lag\gamisone^3\gamistwo\rag_c \lag\gamistwo^2\rag_c 
\lag\gamisone\gamistwo\rag_c + 36 \lag\gamisone^2\gamistwo^2\rag_c 
\lag\gamisone^2\rag_c \lag\gamistwo^2\rag_c \nonumber \\
&& + 72 \lag\gamisone^2\gamistwo^2\rag_c \lag\gamisone\gamistwo\rag_c^2 + 72 
\lag\gamisone\gamistwo\rag_c^2 \lag\gamisone^2\rag_c \lag\gamistwo^2\rag_c 
+ 24 \lag\gamisone\gamistwo\rag_c^4 ,
\label{sigM40M04}
\eeqa
and:
\beqa
\lefteqn{ \sigma^2(M_{22};M_{40}) = \lag\gamisone^6\gamistwo^2\rag_c + 
\lag\gamisone^6\rag_c \lag\gamistwo^2\rag_c + 12 \lag\gamisone^5\gamistwo\rag_c
\lag\gamisone\gamistwo\rag_c + \lag\gamisone^4\gamistwo^2\rag_c 
\lag\gamisone^2\rag_c \left[15+{8\over\rho_1}\right] + 14 \lag\gamisone^4\rag_c
\lag\gamisone^2\gamistwo^2\rag_c } \nonumber \\
&& + 20 \lag\gamisone^3\gamistwo\rag_c^2 + \lag\gamisone^2\gamistwo^2\rag_c 
\lag\gamisone^2\rag_c^2 \left[42+{48\over\rho_1}+{12\over\rho_1^2}\right] 
+ \lag\gamisone^4\rag_c \lag\gamisone^2\rag_c \lag\gamistwo^2\rag_c 
\left[14+{8\over\rho_1}\right] + 28 \lag\gamisone^4\rag_c 
\lag\gamisone\gamistwo\rag_c^2 \nonumber \\
&& + \lag\gamisone^3\gamistwo\rag_c \lag\gamisone\gamistwo\rag_c 
\lag\gamisone^2\rag_c \left[120+{64\over\rho_1}\right] + 
\lag\gamisone^2\rag_c^3 \lag\gamistwo^2\rag_c \left[12+{24\over\rho_1}+
{12\over\rho_1^2}\right] + \lag\gamisone\gamistwo\rag_c^2 
\lag\gamisone^2\rag_c^2 \left[84+{96\over\rho_1}+{24\over\rho_1^2}\right].
\label{sigM22M40}
\eeqa
Finally, we obtain for the estimators $H_{pq}$:
\beq
\sigma^2(H_4) = \sigma^2(M_4) - 12 \lag\gammais^6\rag_c \lag\gammais^2\rag_c
- 12 \lag\gammais^4\rag_c \lag\gammais^2\rag_c^2 \left[11+{8\over\rho}\right]
- 72 \lag\gammais^2\rag_c^4 \left[1+{1\over\rho}\right]^2 ,
\label{sigH4}
\eeq
\beqa
\sigma^2(H_{31}) & = & \sigma^2(M_{31}) - 6 \lag\gamisone^5\gamistwo\rag_c 
\lag\gamisone\gamistwo\rag_c - 6 \lag\gamisone^4\gamistwo^2\rag_c 
\lag\gamisone^2\rag_c - 6 \lag\gamisone^4\rag_c \lag\gamisone^2\rag_c 
\lag\gamistwo^2\rag_c \left[1+{1\over\rho_2}\right] - 21 \lag\gamisone^4\rag_c 
\lag\gamisone\gamistwo\rag_c^2 \nonumber \\ && 
 - \lag\gamisone^3\gamistwo\rag_c \lag\gamisone\gamistwo\rag_c 
\lag\gamisone^2\rag_c \left[78+{36\over\rho_1}\right] 
- \lag\gamisone^2\gamistwo^2\rag_c \lag\gamisone^2\rag_c^2
\left[27+{18\over\rho_1}\right] - \lag\gamisone^2\rag_c^3 
\lag\gamistwo^2\rag_c 9 \left[1+{1\over\rho_1}\right] 
\left[1+{1\over\rho_2}\right] \nonumber \\ 
&& - \lag\gamisone\gamistwo\rag_c^2 \lag\gamisone^2\rag_c^2 
\left[79+{104\over\rho_1}+{34\over\rho_1^2}\right] ,
\label{sigH31}
\eeqa
\beqa
\sigma^2(H_{22}) & = & \sigma^2(M_{22}) - 8 \lag\gamisone^3\gamistwo^3\rag_c 
\lag\gamisone\gamistwo\rag_c - 2 \lag\gamisone^2\gamistwo^4\rag_c 
\lag\gamisone^2\rag_c - 2 \lag\gamisone^4\gamistwo^2\rag_c 
\lag\gamistwo^2\rag_c - 48 \lag\gamisone^2\gamistwo^2\rag_c
\lag\gamisone\gamistwo\rag_c^2 \nonumber \\ 
&& - \lag\gamisone^4\rag_c \lag\gamistwo^2\rag_c^2 - \lag\gamistwo^4\rag_c 
\lag\gamisone^2\rag_c^2 - \lag\gamisone^3\gamistwo\rag_c 
\lag\gamisone\gamistwo\rag_c \lag\gamistwo^2\rag_c \left[32+{16\over\rho_2}
\right] - \lag\gamisone\gamistwo^3\rag_c \lag\gamisone\gamistwo\rag_c
\lag\gamisone^2\rag_c \left[32+{16\over\rho_1}\right] \nonumber \\
&& - \lag\gamisone^2\gamistwo^2\rag_c \lag\gamisone^2\rag_c 
\lag\gamistwo^2\rag_c \left[18+{8\over\rho_1}+{8\over\rho_2}\right] - 16 
\lag\gamisone\gamistwo\rag_c^4 - \lag\gamisone^2\rag_c \lag\gamistwo^2\rag_c 
\lag\gamisone\gamistwo\rag_c^2 \left[52+{32\over\rho_1}+{32\over\rho_2}
+{16\over\rho_1\rho_2}\right] \nonumber \\
&& - \lag\gamisone^2\rag_c^2 \lag\gamistwo^2\rag_c^2 \left[4+{4\over\rho_1}
+{4\over\rho_2}+{2\over\rho_1^2}+{2\over\rho_2^2}\right] ,
\label{sigH22}
\eeqa
\beqa
\sigma^2(H_{31};H_{40}) & = & \sigma^2(M_{31};M_{40}) -3 \lag\gamisone^6\rag_c 
\lag\gamisone\gamistwo\rag_c - 9 \lag\gamisone^5\gamistwo\rag_c 
\lag\gamisone^2\rag_c - \lag\gamisone^3\gamistwo\rag_c \lag\gamisone^2\rag_c^2
\left[66+{48\over\rho_1}\right] \nonumber \\
&& - \lag\gamisone^4\rag_c \lag\gamisone\gamistwo\rag_c \lag\gamisone^2\rag_c 
\left[66+{24\over\rho_1}\right] - \lag\gamisone\gamistwo\rag_c 
\lag\gamisone^2\rag_c^3 \left[72+{108\over\rho_1}+{36\over\rho_1^2}\right] ,
\label{sigH31H40}
\eeqa
\beqa
\lefteqn{\sigma^2(H_{31};H_{04}) = \sigma^2(M_{31};M_{04}) - 3 
\lag\gamisone\gamistwo^5\rag_c \lag\gamisone^2\rag_c - 6
\lag\gamisone^3\gamistwo^3\rag_c \lag\gamistwo^2\rag_c - 3 
\lag\gamisone^2\gamistwo^4\rag_c \lag\gamisone\gamistwo\rag_c 
 - \lag\gamisone^3\gamistwo\rag_c \lag\gamisone^2\rag_c^2 
\left[12+{12\over\rho_2}\right] } \nonumber \\
&& - 12 \lag\gamistwo^4\rag_c \lag\gamisone\gamistwo\rag_c 
\lag\gamisone^2\rag_c  - \lag\gamisone\gamistwo^3\rag_c \lag\gamisone^2\rag_c 
\lag\gamistwo^2\rag_c \left[30+{12\over\rho_2}\right] - 24
\lag\gamisone\gamistwo^3\rag_c \lag\gamisone\gamistwo\rag_c^2 \nonumber \\
&& - 54 \lag\gamisone^2\gamistwo^2\rag_c \lag\gamisone\gamistwo\rag_c 
\lag\gamistwo^2\rag_c - \lag\gamisone\gamistwo\rag_c \lag\gamisone^2\rag_c 
\lag\gamistwo^2\rag_c^2 \left[36+{36\over\rho_2}\right] - 36 
\lag\gamisone\gamistwo\rag_c^3 \lag\gamistwo^2\rag_c ,
\label{sigH31H04}
\eeqa
\beqa
\sigma^2(H_{40};H_{04}) & = & \sigma^2(M_{40};M_{04}) - 6
\lag\gamisone^2\gamistwo^4\rag_c \lag\gamisone^2\rag_c - 6 
\lag\gamisone^4\gamistwo^2\rag_c \lag\gamistwo^2\rag_c - 48 
\lag\gamisone\gamistwo^3\rag_c \lag\gamisone^2\rag_c 
\lag\gamisone\gamistwo\rag_c \nonumber \\
&& - 48 \lag\gamisone^3\gamistwo\rag_c \lag\gamistwo^2\rag_c 
\lag\gamisone\gamistwo\rag_c - 36 \lag\gamisone^2\gamistwo^2\rag_c 
\lag\gamisone^2\rag_c \lag\gamistwo^2\rag_c - 72 
\lag\gamisone\gamistwo\rag_c^2 \lag\gamisone^2\rag_c \lag\gamistwo^2\rag_c ,
\label{sigH40H04}
\eeqa
\beqa
\sigma^2(H_{22};H_{40}) & = & \sigma^2(M_{22};M_{40}) - \lag\gamisone^6\rag_c 
\lag\gamistwo^2\rag_c - 4 \lag\gamisone^5\gamistwo\rag_c
\lag\gamisone\gamistwo\rag_c - 7 \lag\gamisone^4\gamistwo^2\rag_c 
\lag\gamisone^2\rag_c - \lag\gamisone^2\gamistwo^2\rag_c 
\lag\gamisone^2\rag_c^2 \left[30+{24\over\rho_1}\right] \nonumber \\
&& - \lag\gamisone^4\rag_c \lag\gamisone^2\rag_c \lag\gamistwo^2\rag_c 
\left[14+{8\over\rho_1}\right] - 16 \lag\gamisone^4\rag_c 
\lag\gamisone\gamistwo\rag_c^2 - \lag\gamisone^3\gamistwo\rag_c 
\lag\gamisone\gamistwo\rag_c \lag\gamisone^2\rag_c 
\left[72+{16\over\rho_1}\right] \nonumber \\
&& - \lag\gamisone^2\rag_c^3 \lag\gamistwo^2\rag_c \left[12+{24\over\rho_1}+
{12\over\rho_1^2}\right] - \lag\gamisone\gamistwo\rag_c^2 
\lag\gamisone^2\rag_c^2 \left[60+{48\over\rho_1}\right].
\label{sigH22H40}
\eeqa
As for the aperture mass, we can check that the scatter of these estimators
$H_{pq}$ is smaller than for their counterparts $M_{pq}$ and that the galaxy
intrinsic ellipticities only enter their dispersion through the combination
$(1+1/\rho)$.

\end{document}